\documentclass[aps,amsmath,nofootinbib,preprint]{revtex4}
\usepackage{epsfig}
\begin{document}
\vspace*{8cm}

{\bf THERMAL CONDUCTIVITY IN MAGNETIC SUPERCONDUCTORS}
\vspace{25pt}
\newline
\vspace{25pt}
{\bf B. I. Belevtsev$^{\star}$\footnote{E-mail: belevtsev@ilt.kharkov.ua}, 
B. D. Hennings$^{\dagger}$, K. D. D. Rathnayaka$^{\dagger}$, and 
D. G. Naugle$^{\dagger}$\footnote{E-mail: naugle@physics.tamu.edu}}
\newline
$^{\star}${\small{B. Verkin Institute for Low Temperature Physics and 
Engineering, National Academy of Sciences, Kharkov 61103, Ukraine}}
\newline
$^{\dagger}${\small{Physics Department, Texas A\&M University, College
Station, TX 77843-4242, USA}}
\vspace{20pt}



\section{Introduction}
\label{int}
The thermal conductivity is a property that can provide significant 
information about the nature of superconductors.  Even in the early 
studies of high-$T_{\mathrm{C}}$  superconductors before good single crystal 
samples were available, thermal conductivity results indicated the anomalous 
behavior of the normal state transport properties of these oxides and many 
properties of their superconducting state.  (See Uher \cite{uher} for a review 
of early thermal conductivity studies in the high-$T_{\mathrm{C}}$ 
superconductors.)  Recent low temperature thermal conductivity measurements in 
the normal state of an overdoped cuprate superconductor
 Tl$_2$Ba$_2$CuO$_{6+\delta}$ \cite{proust} have demonstrated that the 
fermions which carry heat also carry charge (i.e. the Wiedemann-Franz law is 
fully obeyed, thus providing no evidence for spin-charge separation in 
overdoped high-$T_{\mathrm{C}}$ compounds).  In contrast, this same 
group has observed a breakdown of the Wiedemann-Franz law at low temperatures 
(below $T_0 \approx 0.15$~K) in a different  
high-$T_{\mathrm{C}}$ cuprate (Pr$_{2-x}$Ce$_x$CuO$_{4-\delta}$) near the
optimal doping \cite{hill} which suggests that superconductivity in the 
underdoped region of the phase diagram may indeed be the result of charged 
bosons rather than the Cooper pairing most likely responsible for 
superconductivity in the overdoped regime. It was pointed out in Ref. 
\cite{si}, however, that the experiments of Hill {\it et al.} \cite{hill} 
leave room for possibility that charge-carrying excitations are fermionic, but 
that a subtle transition occurs at  $T_0$. In any case, such experiments help 
to indicate the value of thermal conductivity measurements to probe the
fundamental properties of new materials.
\par
The focus of this work is thermal conductivity measurements in 
systems that appear to exhibit microscopic coexistence of magnetic and 
superconducting order, two types of ordering that are generally antagonistic.  
Although the traditional transport coefficients, Hall effect, thermopower and 
resistivity, provide little or no information in the superconducting phase 
below $T_{\mathrm{C}}$, thermal conductivity measurements can still probe both 
the phonons 
and, at least for a reasonable range of temperature below $T_{\mathrm{C}}$, 
the electron quasiparticles which are the two primary heat carriers in a 
superconductor.  This may be particularly useful for magnetic superconductors 
in the situation where the magnetic ordering temperature $T_{\mathrm{M}}$ is 
less than the superconducting temperature $T_{\mathrm{C}}$  since the electron 
contribution and, perhaps to a lesser extent, the phonon contribution, may 
both be influenced by the magnetic ordering.  At  high temperature the 
thermal conductivity also provides information regarding the scattering 
processes for these carriers and, as will be illustrated, can provide 
indications of subtle high temperature phase transitions.  Thus, thermal 
conductivity can be an important tool for study of new materials over 
the entire temperature range.
\par
In the 1980's, discovery of the microscopic coexistence of both ferromagnetic 
and antiferromagnetic order in the presence of superconducting order in 
various Chevrel phase and rare earth rhodium boride compounds led to a new 
view of these two antagonistic phases.  The experiments generated great 
excitement, and a large number of theoretical predictions of new types of 
order involving spatial modulation of the two very different, mutually 
exclusive order parameters appeared.  Difficulty in synthesis of high quality 
samples of these magnetic superconductors coupled with the excitement of the 
discovery of superconductivity above the boiling point of liquid nitrogen in 
high-$T_{\mathrm{C}}$ oxide superconductors led to greatly reduced activity in 
this field.  (See Maple \cite{maple}, Bulaevskii {\it  et al.} \cite{bulai} 
and Fischer \cite{fischer} for excellent reviews of the state of experiment 
and theory in that era.)  The discovery \cite{mazumdar,nagarajan,cava1,cava2} 
of the family of rare earth nickel borocarbide (RNi$_2$B$_2$C) 
superconductors, which exhibit such a wide variety of interesting phenomena, 
and the widespread availability of good single crystal samples \cite{xu} has 
led to renewed interest in questions regarding the coexistence of 
superconductivity and magnetic order.  This family exhibits superconducting 
order only (R = Y, Lu), heavy fermion behavior (R = Yb), coexistence of 
superconductivity and magnetic order (R = Tm, Er, Ho, Dy) and magnetic order 
only (R = Tb, Gd) within the same crystal structure.  Although we have not 
found any thermal conductivity measurements in the literature for the Chevrel 
and rare earth rhodium boride magnetic superconductors, perhaps due to the 
difficulty of sample preparation, rather complete data now exists for single 
crystals of the rare earth nickel borocarbides.  There are indications of 
coexistence of magnetic  order and superconductivity in a new system, the 2122 
phase of rare earth rutheno-cuprates, also. So far  there appears to be only
one report \cite{brian1} of thermal conductivity, in an oxygen rich 
Eu$_{1.5}$Ce$_{0.5}$RuSr$_2$Cu$_2$O$_{10-\delta}$ compound, but the results of 
this study  suggest  strongly coupled weak 
ferromagnet-antiferromagnet/superconductor-normal metal transitions at a 
remarkably high temperature ($\approx 45$~K).
\par
A brief discussion of the experimental considerations is given in 
section \ref{exp} with a review of the theoretical aspects given in section 
\ref{theor}.  Measurements of the thermal conductivity, together with some 
illustrative data from thermopower and resistivity measurements, for the 
borocarbides and the rutheno-cuprate are discussed in section IV.  Section V 
presents a short conclusion. The reader reasonably familiar with thermal 
conductivity and interested primarily in the results for magnetic 
superconductors may go directly to section \ref{magnsuper}. 

\section{Experimental Techniques}
\label{exp}
We will consider in the following, mainly, the thermal conductivity 
measurements made in the Low Temperature Laboratory  at Texas A\&M University. 
For this reason, we will describe below in sufficient detail the technique 
employed in this lab. This technique is, however, essentially the same as that 
used in many other labs for low temperature measurements of the thermal 
conductivity. A more general and thorough account of available methods of
thermal conductivity measurements can be found in Refs. \cite{parrot,childs}
\par
The thermal conductivity of samples was measured by the steady state 
linear heat flow method. One end of the sample was thermally isolated (a 
known, constant heat input via a resistive heater was applied to that end), 
and the resultant steady state temperature gradient was measured. The 
temperature difference, $\Delta T$, between the two ends of the sample 
(which was typically less than 5\% of the absolute temperature) was measured 
with a Au-0.007 at. \%Fe -- Chromel thermocouple. Verification of the 
temperature difference was provided by a Chromel-Constantan thermocouple for 
most of the temperature range. The temperature differences for each sample and 
the calibration samples (silver foils) were chosen to be approximately the 
same in the different temperature ranges. The sample was completely surrounded 
by a thermal shield held at the temperature of the cold end. Even though heat 
leaks were minimized by design, the actual heat leaks were measured and 
accounted for in the thermal conductivity data. All this suggests 
an absolute accuracy for the thermal conductivity of $\pm 3$ \% with the 
relative precision of $\pm 3$ \% on a specific measurement run. The value of 
the thermal conductivity $\kappa(T)$ is calculated using the
relation
\begin{equation}
\kappa(T)= \frac{L}{A}\frac{{\dot Q}}{\Delta T},
\label{eq1}
\end{equation}
where $L$ and $A$ are length and cross-sectional area of the sample,  
$\dot Q$ is the power input. A detailed discussion
of the thermal-conductivity measurement technique employed can be found in 
\cite{brian2}.
\par
The above-indicated instrumental error in the thermal conductivity 
measurements should be considered as the optimum or, better to say, 
minimum one. Some reservations, therefore, should be added to this. First, it 
is usually more difficult to make a good thermal contact to an oxide sample 
than to a metallic one.  A somewhat greater inaccuracy can be expected 
in the former case. Second, all the thermal conductance 
measurements were made in a vacuum of $1\times 10^{-6}$~Torr or better, so the 
convective heat losses are negligibly small.  The radiative heat losses, due 
to the small temperature difference between the base $T$-end and the 
$\Delta T$-end of the sample mentioned above, should be small except perhaps 
at higher temperatures. As was pointed out in Refs. \cite{uher,parrot}, 
this radiation loss can lead to an overestimate of the measured thermal 
conductivity above $T \approx 150$~K, and the error grows with increasing 
temperature. Measurement of the actual heat leaks without sample, as in the 
present case, provides partial correction for this error. 
Lastly, it is seen from Eq.~(\ref{eq1}) that an appreciable error 
in the absolute determination of the thermal conductivity (as well as in 
supporting measurements of electrical resistivity) is due to inaccuracies in 
the sample dimension measurements. Taking all this into account, it can be said
that the absolute accuracy for borocarbide and rutheno-cuprate samples can be 
worse than $\pm 3$ \%, primarily due to inaccuracies in determination of the 
geometric factors of the somewhat irregular single crystal samples and 
contact resistance for the pressed powder rutheno-cuprate samples. 
\par
The electrical resistivity, $\rho(T)$, was measured for all of the samples 
studied (except the rutheno-cuprate) using the standard four-terminal method. 
These measurements are helpful and even imperative in many respects. First of 
all, it gives an indication of the quality of the sample. Second, $\rho(T)$ 
curves have in general some peculiarities at magnetic transitions  and show 
the resistive transition at the superconducting transition temperature, 
$T_{\mathrm{C}}$. Last but not least, $\rho(T)$ data allow use of the 
Wiedemann-Franz law to estimate the fraction of the total thermal 
conductivity, that is due to the charge-carrier contribution, and, hence, the 
remainder can thought to be due to phonons and magnetic effects, if any.  
In the case of rare-earth nickel borocarbides, both the resistivity and 
thermal conductivity  were measured in the $ab$-plane.

\section{Theoretical Concepts} 
\label{theor}
For consideration and discussion of available thermal conductivity data for 
the magnetic superconductors, we will refer to the well established 
physical mechanisms for heat transfer in normal metals and superconductors, 
outlined in well known books or papers 
\cite{uher,parrot,ziman,blatt,ashcroft,berman,geilik,brt,gk,tewordt,gk2,rick,ginsberg}. With this we will try to 
understand and explain what is observed in magnetic superconductors and make 
some conclusions about the properties of heat carriers in them. Since the main
heat carriers in conducting solids are charge carriers (electrons or holes) 
and lattice thermal excitations (phonons), the thermal conductivity behavior 
reflects  properties of these quasiparticles and an interaction between them. 
In magnetic states the magnetic excitations (i.e. magnons) can also 
participate in the heat transport. Hence the thermal conductivity measurements 
can provide an insight into  the nature of the superconducting and magnetic 
states of magnetic superconductors. 
\par
The main (and sometimes rather difficult) task in an analysis of the thermal 
conductivity data for some particular material is to separate the 
contributions from the two main type of the heat carriers -- electrons and 
phonons. This requires, primarily, a knowledge about an expected temperature 
behavior of these contributions (which are quite different for electron and 
phonon heat carriers). Crystal-lattice disorder can strongly effect both 
the electron and phonon parts of thermal conductivity and, therefore, must 
generally be considered. Generally speaking, it is important to find out what 
scattering mechanism limits the heat transport for given channel 
(electrons/phonons) at given conditions, which are determined by the phase or 
state of a sample, its 
temperature, the magnitude of applied magnetic field and other circumstances. 
\par
In the simplest way the thermal conductivity of any type of heat carriers 
can be presented by the equation \cite{parrot,ziman,ashcroft,berman}
\begin{equation}
\kappa(T)= \frac{1}{3}C_{\mathrm{v}}{\bar v}l=
\frac{1}{3}C_{\mathrm{v}}{\bar v}^{2}\tau ,
\label{eq2}
\end{equation}
where $C_{\mathrm{v}}$ is the heat capacity per unit volume at constant volume,
${\bar v}$ is the average velocity of the heat carriers, $l$ is the mean free 
path of the heat carriers (that is the average distance they travel between the
collisions with any obstacles to heat transport), and $\tau$ is the 
corresponding relaxation time. Since the quantity $D=(1/3){\bar v}l$ is just a 
diffusion coefficient of the heat carriers, it can be said that the thermal 
conductivity is determined by the heat capacity and diffusivity of the 
heat carriers.  In line with this, its temperature dependence, $\kappa (T)$, 
is determined by those of the heat capacity and the mean free path  of the 
heat carriers, since ${\bar v}$ is essentially temperature independent 
(although it can undergo a change at some phase transitions).
\par
Eq.~(\ref{eq2}) is applicable for both electron and phonon heat 
transport in solids. Generally, the heat transport via electrons and phonons 
occurs in parallel. It can be written then
\begin{equation}
\kappa= \kappa_{\mathrm{p}}+ \kappa_{\mathrm{e}},
\label{eq3}
\end{equation}
where $\kappa_{\mathrm{p}}$ and $\kappa_{\mathrm{e}}$ present the 
contributions from the phonons and free charge carriers, respectively. Below we
consider these contributions separately. The phonon heat transport takes place 
in any solid; whereas, the charge carrier contribution depends on the carrier 
density, and is, therefore, negligible in insulators. It should be noted as
well that different types of relaxation processes, which act simultaneously, 
 can limit the thermal and electrical conductivities. As a quite good 
approximation, Matthiessen's rule \cite{ziman,blatt}, is applied in this 
case to describe the combined effect of these processes:
\begin{equation}
\frac{1}{\tau_{\mathrm{eff}}}=\sum_{i} \frac{1}{\tau_{\mathrm{s}}^{i}},
\label{eq4}
\end{equation}
where $\tau_{\mathrm{eff}}$ is the effective (or total) relaxation time, and 
the times $\tau_{\mathrm{s}}^{i}$ correspond to different relaxation 
processes. Since  the mean free path is proportional to the relaxation time 
($l = {\bar v}\tau$), the same relation can be written for the effective 
mean free path, $l_{\mathrm{eff}}$. According to the Matthiessen's rule, the 
electrical resistivity can be presented as a sum of the partial resistivities,
corresponding to different processes of the electron scattering. By analogy, 
it is convenient in many cases to consider the thermal resistivity, 
$W=1/\kappa$, \cite{ziman,berman} as a sum of partial resitivities
\begin{equation}
1/\kappa_{\mathrm{eff}} = W = \sum_{i} W_{i}
\label{eq5}
\end{equation}

\subsection{Phonon thermal conductivity in non-metallic crystals}
\label{phonon}
For phonon thermal conductivity, one can use in Eq.~(\ref{eq2}) 
the lattice specific heat $C_{\mathrm{p}}$, and ${\bar v}$ may be 
taken as an average velocity of sound, $v_{\mathrm{s}}$. As a good 
approximation, the Debye model for the lattice specific heat 
\cite{ziman,ashcroft} can be used. According to it,
\begin{subequations}
\label{C_a}
\begin{eqnarray}
C_{\mathrm{p}} & \propto & T^3  \hspace{2.5cm}  
(T\ll\Theta_{\mathrm{D}}), \label{C_va} \\
C_{\mathrm{p}} & = & 3Nk_{\mathrm{B}} \hspace{2.1cm} 
(T >\Theta_{\mathrm{D}}), 
\label{C_vb}
\end{eqnarray}
\end{subequations}
where $N$ is the ion density. The main problem is 
to take into account the most important  mechanisms of phonon relaxation 
in crystal solids. In doing so it is convenient to operate with the 
thermal resistivity, which can be presented in Matthiessen's approximation as
\begin{equation}
1/\kappa_{\mathrm{p}} = W_{\mathrm{p}} = W_{\mathrm{pp}}+
W_{\mathrm{pi}}+W_{\mathrm{pe}} =
\frac{3}{C_{\mathrm{p}}v_{\mathrm{s}}^{2}}\left (
{\tau_{\mathrm{pp}}}^{-1} + {\tau_{\mathrm{pi}}}^{-1} +
{\tau_{\mathrm{pe}}}^{-1}\right ),
\label{eq7}
\end{equation}
where subscripts {\tt pp}, {\tt pi} and {\tt pe} indicate the phonon-phonon, 
phonon-imperfection and phonon-electron interactions, respectively.
\par
The phonon-phonon and phonon-imperfection interactions, corresponding to the 
first two terms in Eq.~(\ref{eq7}), occur in any solid; whereas, the 
phonon-electron interaction can be important only in solids with fairly high 
charge-carrier density. Let us consider, at first, a rather perfect 
non-metallic crystal, where the main contribution to the thermal conductivity 
comes from the phonon-phonon interaction \cite{ziman,ashcroft,berman,parrot}. 
These processes
are those in which two phonons can combine to give a third, and vice versa (so 
called, three-phonon processes). They are restricted by the following 
selection rules:  
\begin{subequations}
\label{select}
\begin{eqnarray}
\hbar\omega_1 + \hbar\omega_2   = \hbar\omega_3,  \label{sela} \\
{\mathbf{q}}_{1}+ {\mathbf{q}}_{2}  =  {\mathbf{q}}_{3} + {\mathbf{g}}, 
\label{selb}
\end{eqnarray}
\end{subequations}
where the first condition is conservation of energy at the interaction. In the
second condition for wave vectors of interacting phonons, $\mathbf{g}$ is a 
reciprocal lattice vector. The case 
$\mathbf{g}=0$ corresponds to the conservation of momentum (or wave vector). 
These processes are called {\it Normal} processes or $N$-processes. The case 
$\mathbf{g} \neq 0$ is determined by the interference condition for wave
vectors (the lattice acts in that event as a diffraction grating). These 
processes are called {\it Umklapp} processes or $U$-processes. 
\par
Since $N$-processes are energy and momentum conserving they do not contribute 
to the thermal resistivity (see Refs. \cite{ziman, ashcroft,berman,parrot} for 
thorough explanations). These processes can, however, affect the thermal 
conductivity indirectly, enhancing or modifying the effect of other scattering 
mechanisms, i.e. the phonon-imperfection scattering (see discussion of some 
examples in Refs. \cite{parrot,berman}).  In contrast to $N$-processes, the 
large changes in crystal momentum in $U$-processes make it possible for 
them to be an effective source of thermal resistivity. The probability of 
$U$-processes is, however, temperature dependent, so that they play a 
dominant part at high temperatures ($T\agt \Theta_{\mathrm{D}}$), but are 
of little importance at low temperatures ($T \ll \Theta_{\mathrm{D}}$), where  
only $N$-processes can occur at an appreciable rate. The reason is that at any 
temperature $T$, only the so called, thermal or dominant phonons with energy 
$\hbar \omega \simeq k_{B}T$ are present in an appreciable number and can, 
therefore, be involved significantly in different interactions with other 
quasiparticles (phonons included). The wave length, $\lambda_{\mathrm{p}}$, 
and  modulus, $q_{\mathrm{T}}$, of the wave vector of thermal  
phonons are given by
\begin{equation}
\lambda_{\mathrm{p}} = 2\pi \frac{\hbar v_{\mathrm{s}}}{k_{\mathrm{B}}T},
\label{lambda}
\end{equation}
and 
\begin{equation}
q_{\mathrm{T}} = \frac{k_{\mathrm{B}}T}{\hbar v_{\mathrm{s}}} = 
\left (\frac{T}{\Theta_{\mathrm{D}}}\right )q_{\mathrm{D}},
\label{q}
\end{equation}
where $q_{\mathrm{D}}$ is the Debye wave vector. 
\par
It can be seen from Eqs.~(\ref{select}) that to ensure an $U$-process,  some 
of the participating phonons should have a large enough wave vector,
at least $q \simeq (1/2) g$ according to \cite{ziman}, (i. e., comparable 
with $q_{\mathrm{D}}$) and high enough energy (comparable with 
$k_{\mathrm{B}}\Theta_{\mathrm{D}}$).  Thus, the $U$-processes can have an 
appreciable rate only 
at temperatures in the vicinity of or above $\Theta_{\mathrm{D}}$. 
\par
Now consider the temperature dependence of the thermal conductivity, 
$\kappa (T)$, of a rather perfect non-metallic crystal.  At sufficiently high 
temperatures ($T > \Theta_{\mathrm{D}}$), the specific heat of the crystal 
should be constant [Eq.~(\ref{C_vb})], and, therefore, the temperature 
dependence $\kappa (T)$ is determined solely by that of the phonon-phonon 
scattering rate, $\tau_{\mathrm{pp}}^{-1}$. The dominant contribution to the 
thermal resistivity in this temperature range is from $U$-processes. The 
rate,  $\tau_{\mathrm{pp}}^{-1}$, should increase with temperature, since  the 
total number of phonons (which are scatterers to other phonons which carry 
heat) is proportional to $T$ in this temperature range. It is expected 
\cite{ziman,ashcroft,berman,parrot} that 
$\tau_{\mathrm{pp}}^{-1}\propto T^{n}$ with $n=1$ for the three-phonon 
processes, which is to say that $\kappa_{\mathrm{p}} \propto 1/T$. More 
elaborate theoretical calculations \cite{ziman,berman} give 
\begin{equation}
\kappa_{\mathrm{p}} \propto 
\frac{aM_{\mathrm{a}}\Theta_{\mathrm{D}}^{3}}{T\gamma_{\mathrm{G}}^{2}},
\label{hight}
\end{equation} 
where $a^3$ gives the volume occupied by one atom, $M_{\mathrm{a}}$ is the
atomic weight, $\gamma_{\mathrm{G}}$ is Gr\"{u}neisen constant. The $1/T$ law 
can be considered as a good approximation for the lattice thermal conductivity 
of fairly perfect crystals at $T>\Theta_{\mathrm{D}}$ \cite{berman}. 
\par
For $T<\Theta_{\mathrm{D}}$, the probability of $U$-processes  drops sharply
as temperature decreases. In this temperature range theoretical calculations
give $\tau_{\mathrm{pp}}\propto \exp (\Theta_{\mathrm{D}}/bT)$ and 
\begin{equation}
\kappa_{\mathrm{p}} \propto T^{x}\exp (\Theta_{\mathrm{D}}/bT),
\label{low}
\end{equation} 
with $x$ and $b$ both of the order of unity \cite{ashcroft,ziman,berman}.
In this case the relaxation time $\tau_{\mathrm{pp}}$ and thermal conductivity 
increase exponentially with decreasing temperature. This increase lasts 
until the phonon-phonon mean free path, 
$l_{\mathrm{pp}} = v_{\mathrm{s}}\tau_{\mathrm{pp}}$, becomes comparable with 
that of the phonon-imperfection scattering or even with dimensions of the 
sample (boundary scattering). In this case the effective time of phonon 
relaxation becomes temperature independent, and the temperature dependence of 
$\kappa_{\mathrm{p}}$ will be determined by that of the specific heat, which 
reduces with temperature as $T^3$ for $T\ll \Theta_{\mathrm{D}}$. 
\par
In summary, the $\kappa_{\mathrm{p}} (T)$ dependence for fairly perfect 
non-metallic crystals behaves as follows. At high temperatures 
$T>\Theta_{\mathrm{D}}$, it is proportional to  $1/T$ ($1/T$ law). As 
temperature decreases below $\Theta_{\mathrm{D}}$, $\kappa_{\mathrm{p}}(T)$ 
increases exponentially according to Eq.~(\ref{low}) reaching a maximum at a 
temperature, at which the phonon mean free path begins to
be determined by imperfections or outer boundaries of the sample. With 
decreasing temperature, the mean free path becomes temperature independent, 
and, therefore, the temperature dependence of $\kappa_{\mathrm{p}}$ is 
determined by that of the specific heat, which is proportional to $T^3$ for 
$T\ll\Theta_{\mathrm{D}}$. 
\par 
A clear manifestation of this exponential law [Eq.~(\ref{low})] is confined to 
range $1/30 < T/\Theta_{\mathrm{D}} < 1/10$ \cite{berman}. This sets an upper 
temperature limit for the position of the thermal conductivity peak. The peak 
can be seen, however, only in rather pure and perfect non-metallic crystals.  
Crystal-lattice imperfections (lattice defects) are effective phonon 
scatterers. They cause an extra thermal resistance which can even suppress 
the thermal conductivity peak completely. Although rather considerable 
theoretical and experimental efforts were made to resolve this problem (see 
\cite{ziman,berman} and references therein), the understanding level 
achieved only allows  some general discussion and speculation regarding  
analysis of imperfection effects in experimental thermal conductivity data. 
\par
The important feature of lattice defects is whether their linear dimensions 
are larger or smaller than the phonon wavelength. The first type of defects 
include, for example, external or internal (grain) boundaries,  dislocations. 
stacking faults. The second 
type is represented mainly by point defects or small precipitate particles due 
to phase inhomogeneity. It is obvious that some defects can change their type  
with changing temperature since the wavelength, $\lambda_{\mathrm{p}}$, of 
dominant phonons depends  on temperature as $1/T$ [Eq.~(\ref{lambda})]. It can 
encompass hundreds of interatomic distances for $T\ll\Theta_{\mathrm{D}}$; 
whereas, it is of the order of the interatomic distance at 
$T\geq\Theta_{\mathrm{D}}$. This is in sharp contrast with the wavelength of
electrons in good metals, which is always of the order of the interatomic 
distance, independent of temperature. 
\par
Among other imperfections, the point defects are considered as the most 
important source of thermal resistance. These are impurity atoms, 
isotopes, lattice vacancies and interstitial atoms. The point defects 
introduce small perturbations of mass, force constant and nearest-neighbor 
distance, causing elastic phonon scattering. Their size is of order of the
interatomic distance. The known theoretical results \cite{ziman,berman} give, 
in most cases, only general ideas about the influence of point defects 
on the thermal conductivity. At low temperature $T\ll\Theta_{\mathrm{D}}$ 
(when $\lambda_{\mathrm{p}}$ is well over the size of point defects)  the
mechanism of phonon-point defect relaxation is thought to be similar to that
of Rayleigh scattering, for which the relaxation rate is proportional to 
$\omega{^4}$ \cite{ziman,berman}. Since $\omega\approx k_{\mathrm{B}}T/\hbar$ 
for dominant phonons, the relation, $\tau_{\mathrm{pi}}^{-1}\propto T^4$, 
should hold for the rate of phonon-point defect relaxation. In this case, for 
the low temperature range where phonon specific heat is proportional to $T^3$, 
the thermal conductivity should be proportional to $1/T$. If phonon-point 
defect scattering is dominant 
at low temperatures, the exponential rise in thermal conductivity due to 
$U$-processes [Eq.~(\ref{low})] will be suppressed.  At temperatures so low 
that the phonon mean free path, 
$l_{\mathrm{pi}} = v_{\mathrm{s}}\tau_{\mathrm{pi}}$, 
associated with phonon-point defect scattering, becomes comparable with 
the sample dimensions and, hence, temperature independent, the 
thermal conductivity goes to zero with decreasing temperature according to 
relation $\kappa_{\mathrm{p}}\propto T^3$. 
\par
As temperature increases, the phonon wavelength, $\lambda_{\mathrm{p}}$, 
decreases, approaching an interatomic distance. This causes the rate of 
phonon-point defect relaxation to become temperature independent. At high 
enough temperature this is true for phonon relaxation by any lattice defects, 
regardless of the temperature scattering law for the defects at low 
temperature \cite{berman}. It follows from the aforesaid that the 
phonon-point defect contribution to thermal resistivity can be appreciable 
at high temperatures ($T\agt\Theta_{\mathrm{D}}$), where it depends only 
slightly on temperature, while at low enough temperatures the phonon 
wavelength becomes so long that scattering by point defects (and by most of 
the other lattice defects) becomes negligible. It should be
recalled that the electron wavelength in good metals is always of the order of 
the interatomic distance, and electron-imperfection scattering is temperature 
independent. For this reason, imperfections remain the only source 
of electron scattering for temperatures so low that the electron-phonon 
relaxation processes are essentially frozen out. This important difference in 
the behaviors of electrons and phonons, as heat carriers, should be taken into
account in consideration of the electron and phonon contributions to thermal 
resistance at low temperatures. 
\par
In conductors with a fairly high free electron concentration, the 
phonon-electron scattering [see Eq.~(\ref{eq7})] can give an appreciable 
contribution to thermal resistance. This problem is, however, so closely 
connected with electrical and thermal conductivities of electrons that it is
appropriate to consider it after the main concepts of the electronic 
thermal conductivity will be discussed in the next section. 

\subsection{Electronic thermal conductivity}
\label{subetc}
For considerations of electronic thermal conductivity, the general 
expression given by Eq.~(\ref{eq2}) is used, taking $C_{\mathrm{v}}$ to be the 
electronic specific heat $C_{\mathrm{e}}$ and ${\bar v}$ is the Fermi velocity 
$v_{\mathrm{F}}$, i. e. 
\begin{equation}
\kappa_{\mathrm{e}}(T)= \frac{1}{3}C_{\mathrm{e}}v_{\mathrm{F}}l =
\frac{1}{3}C_{\mathrm{e}}v_{\mathrm{F}}^{2}\tau .
\label{etc}
\end{equation} 
The electronic specific heat in the free-electron model 
\cite{ziman, blatt,ashcroft} is given by
\begin{equation}
C_{\mathrm{e}} = \frac{\pi^{2}}{2} \left (
\frac{k_{\mathrm{B}}T}{{\cal E}_{\mathrm{F}}}\right )
n_{\mathrm{e}}k_{\mathrm{B}} = \gamma T,
\label{esh}
\end{equation} 
where ${\cal E}_{\mathrm{F}}$ is the Fermi energy, $n_{\mathrm{e}}$ is 
electronic density, and $\gamma$ is the Sommerfeld coefficient. The linear 
dependence on temperature is quite general, however, and holds for any Fermi 
liquid, independent of the free-electron model.
\par
It is of fundamental importance that the same scatterers (phonons and lattice 
defects) determine the resistance to both the charge and the heat transport of 
electrons. The electronic conductivity, $\sigma$,  in the frame of 
free-electron model is given by \cite{ziman, blatt,ashcroft} 
\begin{equation}
\sigma = \frac{n_{\mathrm{e}}e^{2}\tau}{m}.
\label{econd}
\end{equation} 
When expressions for electronic thermal and electrical conductivities 
[Eqs.~(\ref{etc}) and Eq.~(\ref{econd})] are compared to one another, it is 
apparent (using Eq.~({\ref{esh}) for electronic specific heat as well) that 
$\kappa_{\mathrm{e}}/(\sigma T)$} should be constant, if the electron 
relaxation times (the total relaxation time, $\tau$) are equal for both kinds 
of electron transport. This simple relation is the famous Wiedemann-Franz (WF) 
law. In the free-electron model this law is expressed as 
\cite{ziman,ashcroft,blatt}
\begin{equation}
\frac{\kappa_{\mathrm{e}}}{\sigma T}= L_{0},  
\label{wf}
\end{equation}
where $L_{0}=\left (\pi^2/3 \right )\left ( k_{\mathrm{B}}/e\right )^{2}$ is 
called the Lorenz number. 
\par
The WF law forms the basis for separation of the electron and phonon 
contributions to  thermal resistance during analysis of experimental data on 
thermal conductivity. This law appears, however, to be true only at the lowest 
temperatures, where electron-imperfection scattering is dominant, and at high 
temperatures ($T/\Theta_{\mathrm{D}}\agt 1$). In the intermediate temperature 
range,  a significant violation of the WF law takes place. The reason is that 
the electrical and thermal conductivities are determined by different types of 
electron-phonon collisions in this intermediate temperature range. To make 
this point more clear, let us write in the Matthiessen's approximation the 
more detailed expressions for the thermal and electrical resistivities:
\begin{equation}
1/\kappa_{\mathrm{e}} = W_{\mathrm{e}} = W_{\mathrm{ep}} + W_{\mathrm{ei}} =
\frac{3}{C_{\mathrm{e}}v_{\mathrm{F}}^{2}}\left [{\tau_{\mathrm{ei}}}^{-1} +
{\tau_{\mathrm{ep(\kappa)}}}^{-1}(T)
\right ],
\label{etres}
\end{equation}
\begin{equation}
\rho = \rho_{\mathrm{ep}} + \rho_{\mathrm{ei}} =
\frac{m}{n_{\mathrm{e}}e^{2}}\left [{\tau_{\mathrm{ei}}}^{-1} +
{\tau_{\mathrm{ep(\sigma)}}}^{-1}(T)
\right ],
\label{eeres}
\end{equation}
where subscript {\tt ei} indicates the electron-imperfection interaction 
(which is temperature independent), and subcripts {\tt ep}($\kappa$) and 
{\tt ep}($\sigma$) indicate electron-phonon interactions, crucial for 
thermal and electrical conductivities, respectively. Of course, some other 
sources of electron scattering can be indicated as well (for example, 
electron-electron scattering or the spin disorder in paramagnetic, 
ferromagnetic or antiferromagnetic states), but the two mentioned above are 
considered as the most significant.
\par  
The times $\tau_{\mathrm{ep(\sigma)}}$ and $\tau_{\mathrm{ep(\kappa)}}$ 
are characterized by the following temperature behaviors 
\cite{ziman,blatt,ashcroft}: 
\begin{equation}
\tau_{\mathrm{ep(\sigma)}}^{-1}(T) \propto
\begin{cases}
T^5& \text{if $T\ll\Theta_{\mathrm{D}}$},\\
T& \text{if $T\gtrsim \Theta_{\mathrm{D}}$};
\end{cases}
\label{taur}
\end{equation}

\begin{equation}
\tau_{\mathrm{ep(\kappa)}}^{-1}(T) \propto
\begin{cases}
T^3& \text{if $T\ll\Theta_{\mathrm{D}}$},\\
T& \text{if $T\gtrsim \Theta_{\mathrm{D}}$}.
\end{cases}
\label{taut}
\end{equation}
It is seen that temperature dependences of these times are the same near or 
above $\Theta_{\mathrm{D}}$, but are quite different below it. Consider 
briefly the sources of this diversity. When an electron with wave vector 
$\mathbf{k}$ is scattered by a phonon into the state $\mathbf{k}^{'}$, the 
electron either absorbs or emits a phonon of wave vector $\mathbf{q}$. 
These collisions are determined by the following rules 
\cite{ziman,blatt,ashcroft}:
\begin{subequations}
\label{sele}
\begin{eqnarray}
\mathbf{k}^{'} = \mathbf{k} \pm \mathbf{q} + \mathbf{g}, \label{ela} \\ 
{\cal E}_{{\mathbf{k}^{'}}}  = {\cal E}_{\mathbf{k}} \pm 
\hbar\omega_{\mathbf{q}}.
\label{elb}
\end{eqnarray}
\end{subequations}  
The first rule maintains that momentum is conserved, up to the addition of any
arbirary vector of the reciprocal lattice. In the same way as for the 
phonon-phonon scattering (see Sec. \ref{phonon}), the cases 
$\mathbf{g} = 0$ and $\mathbf{g} \neq 0$ correspond to $N$- and $U$-processes,
respectively. It should be noted that the $U$-processes of the electron-phonon 
interaction are very improbable at low temperature, but they can increase 
somewhat the electronic electrical and thermal resistivities at fairly high 
temperatures (above $\approx 0.2$~$ \Theta_{\mathrm{D}}$). The second rule  
is the requirement for conservation of energy in the interaction.
\par
The maximum phonon energy is about $k_{\mathrm{B}}\Theta_{\mathrm{D}}$, which  
is much less than the electron energy, ${\cal E}_{\mathrm{F}}$. For this 
reason, the electron-phonon collisions could be considered as quasi-elastic 
ones.  For electronic thermal conduction, however, it is more important 
to compare the change in electron energy in a collision  
($\Delta {\cal E}_{\mathbf{q}} = 
|{\cal E}_{{\mathbf{k}}} - {\cal E}_{\mathbf{k}^{'}}|$) with 
$k_{\mathrm{B}}T$ (since electrons in metals can change their energy only 
in narrow band near the Fermi level with a width about $k_{\mathrm{B}}T$). 
Following this criterion, the collisions with 
$\Delta {\cal E}_{\mathbf{q}} \simeq k_{\mathrm{B}}T$ are considered as 
inelastic ones while those with 
$\Delta {\cal E}_{\mathbf{q}} < k_{\mathrm{B}}T$, as elastic 
\cite{ziman,blatt,ashcroft}. 
These collisions are quite effective for electronic thermal resistance. At low 
temperatures, $T\ll \Theta_{\mathrm{D}}$, they are inelastic (since the energy 
of the dominant phonon is about $k_{\mathrm{B}}T$ in this temperature range) 
with a collision rate (or an inverse relaxation time, $1/\tau_{\mathrm{ep}}$) 
proportional to $T^3$ \cite{ziman,ashcroft}. This corresponds to the low 
temperature behavior of time 
$\tau_{\mathrm{ep(\kappa)}}$ in Eq.~(\ref{taut}), governing the electronic 
thermal resistance.  
\par
At low temperatures,  $T\ll \Theta_{\mathrm{D}}$, a single collision, 
which must be an $N$-process according to the selection of  Eq.~({\ref{sele}), 
is, however, ineffective in providing  electrical resistance. The reason is 
that the electron direction of motion can be changed only slightly in this 
$N$-process.  The angle 
between vectors $\mathbf{k}$ and $\mathbf{k}^{'}$ (the angle of scattering) is 
$\theta \simeq q_{\mathrm{T}}/k_{\mathrm{F}} = T/\Theta_{\mathrm{D}}$, that 
is, very small.  An  electron must endure many of such collisions to gain a 
large scattering angle, and the probability of that event is proportional 
to $(T/\Theta_{\mathrm{D}})^2$ \cite{ziman,blatt,ashcroft}.  For this reason,
the effective rate of the electron-phonon relaxation for electrical resistance 
is $\tau_{\mathrm{ep(\sigma)}}^{-1}(T) \propto 
T^{3}(T/\Theta_{\mathrm{D}})^2 \propto T^{5}$, which was already indicated in
Eq.~(\ref{taur}). 
\par
At high temperatures, $T>\Theta_{\mathrm{D}}$, any electron-phonon collision 
is large-angle and elastic. Since the total number of phonons is proportional 
to $T$ at this temperature range, the relation 
$\tau_{\mathrm{ep}}^{-1}\propto T$ holds for both electron charge and heat 
transport. 
\par
We can survey now a temperature dependence of the electronic thermal 
conductivity, taking into account Eqs.~(\ref{etc}), (\ref{esh}) and 
(\ref{etres}). At temperatures so low that the elastic and temperature 
independent electron-imperfection scattering is dominant, the thermal 
conductivity behaves as $\kappa_{\mathrm{e}}(T) \propto T$. In this 
temperature range the WF law [Eq.~(\ref{wf})] holds.  For higher temperatures, 
the electron-phonon scattering becomes dominant. At 
 $T\ll\Theta_{\mathrm{D}}$, this scattering is inelastic with rate,
$\tau_{\mathrm{ep(\kappa)}}^{-1} \propto T^3$, relevant for heat transport, 
and the rate $\tau_{\mathrm{ep(\sigma)}}^{-1} \propto T^5$, relevant for 
charge transport.  This leads to the relation $\kappa_{\mathrm{e}}(T) 
\propto T^{-2}$. Due to different electron-phonon relaxation rates, 
relevant for charge and heat transport, the WF law is not obeyed in this
temperature range, so that the ratio $\kappa_{\mathrm{e}}/(\sigma T)$ is not
constant, but is proportional to $T^2$.  
At high temperatures, $T>\Theta_{\mathrm{D}}$, electron-phonon collisions 
are elastic with rate $\tau_{\mathrm{ep}}^{-1}\propto T$, which gives 
relations, $\rho \propto T$, for electrical resistivity and,
$\kappa_{\mathrm{e}}(T) = const$,  for thermal conductivity. The WF law holds 
in this temperature range. For fairly perfect metals with high electron 
density, the electron contribution dominates in thermal conductivity, so that
the outlined features of  $\kappa_{\mathrm{e}}(T)$ dependence are immediately 
evident from measured $\kappa (T)$ curves. An example of such behavior is 
presented in Fig. \ref{ag} for a well annealed silver foil \cite{brian2}.
\begin{figure}[tbh]
\vspace{-20pt}
\centerline{\epsfig{file=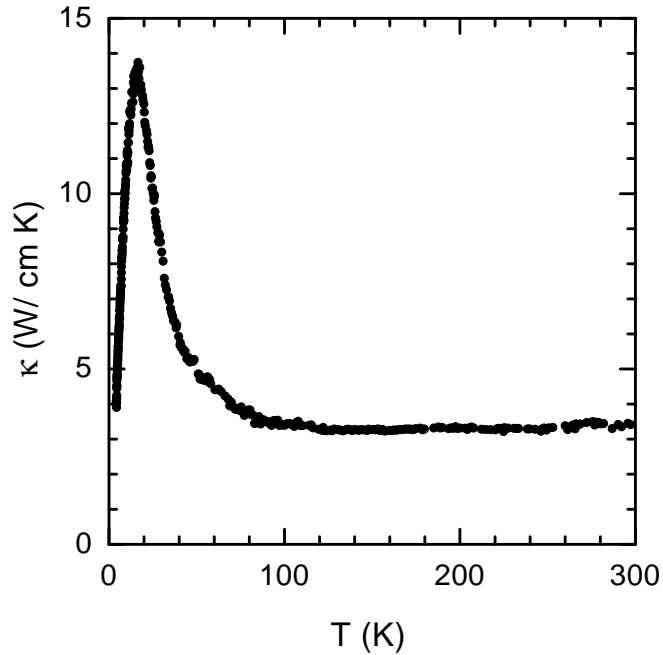,width=10cm}}
\vspace{-20pt}
\caption{{\small {\sl $\kappa_{\mathrm{e}}(T)$ dependence of annealed silver 
foil.}}} 
\vspace{10pt}
\label{ag}
\end{figure}
\par 
It is seen that a concurrence of the electron-imperfection and electron-phonon 
interactions gives rise to a maximum (a peak) in the $\kappa_{\mathrm{e}}(T)$ 
dependence. This peak can be very sharp for fairly perfect and pure metals, 
where the peak position, $T_{\mathrm{max}}$, is usually below 0.1 
$\Theta_{\mathrm{D}}$. Lattice 
defects induce the following: (i) an increase in slope of the linear part of 
$\kappa_{\mathrm{e}}(T)$ dependence; (ii) an increase in $T_{\mathrm{max}}$; 
and (iii) a decrease in the peak height up to a total smearing of it for 
strong enough lattice disorder. Some clear demonstrations of this imperfection 
effect
for typical metals can be found in \cite{parrot,childs,blatt,berman,white}. It 
is apparent also [see Eq.~(\ref{etres})] that in ``bad'' metals with strong 
electron-imperfection scattering the WF law is obeyed better (even in the 
intermediate temperature range where inelastic electron-phonon scatering 
occurs) than in pure metals with a perfect crystal lattice.  

\subsection{Effects of phonon-electron interactions in thermal conductivity}
\label{perel}
The phonon-electron scattering can limit the phonon thermal conductivity (this 
point was briefly mentioned at the end of Sec.~\ref{phonon}). This interaction 
leads to absorption or emission of phonons by electrons and is restricted by 
the above-mentioned rules for electron-phonon interaction [Eq.~(\ref{sele})]. 
The thermal resistivity, determined by these processes, is labelled 
$W_{\mathrm{pe}}$ in the general expression (\ref{eq7}}) for phonon thermal 
resistivity with an associated relaxation time $\tau_{\mathrm{pe}}$. The known 
theoretical examinations \cite{ziman,parrot,berman} have shown that 
$\tau_{\mathrm{pe}}^{-1}(T) \propto T$ if $T\ll\Theta_{\mathrm{D}}$, and is 
constant if $T > \Theta_{\mathrm{D}}$. This leads [taking into account 
Eqs.~(\ref{C_a}) and (\ref{eq7})] to 
\begin{equation}
W_{\mathrm{pe}} \propto
\begin{cases}
T^{-2}& \text{if $T\ll\Theta_{\mathrm{D}}$},\\
\mathrm{const}& \text{if $T > \Theta_{\mathrm{D}}$}.
\end{cases}
\label{kappe}
\end{equation}
Equations (\ref{kappe}) actually results from the  more general relation 
\cite{ziman}
\begin{equation}
\frac{W_{\mathrm{pe}}T}{\rho} \leq 
\left (\frac{e}{k_{\mathrm{B}}}\right )^{2}n_{\mathrm{a}}^{2} 
\left (\frac{3Nk_{\mathrm{B}}}{C_{\mathrm{p}}}\right )^{2},
\label{pei}
\end{equation}
where $n_{\mathrm{a}}$ is the number of conduction electrons per atom. Besides 
the electron-phonon $N$-processes, this equation takes into account the 
$U$-processes as well, and that causes an appearance of the nonequality sign. 
It is easy to verify that in the case of the equality sign (that is, 
ignoring the $U$-processes) the relationships (\ref{kappe}) can be derived 
from Eq.~(\ref{pei}) for low ($T\ll\Theta_{\mathrm{D}}$) as well as for high 
($T>\Theta_{\mathrm{D}}$) temperatures. In the former case, the relationships 
$\rho \propto T^{5}$ and $C_{\mathrm{p}} \propto T^3$ should be taken, 
and in the latter case, $\rho \propto T$ and 
$C_{\mathrm{p}} = 3Nk_{\mathrm{B}}$.
\par
Equation (\ref{pei}) has a doubtless similarity with the WF law presented by 
Eq.~(\ref{wf}). At high temperatures ($C_{\mathrm{p}} = 3Nk_{\mathrm{B}}$)
it is possible to rewrite it as
\begin{equation}
\frac{\rho}{W_{\mathrm{pe}}T} = \frac{\kappa_{\mathrm{pe}}}{\sigma T} \geq 
\left (\frac{k_{\mathrm{B}}}{e}\right )^{2} 
\frac{1}{n_{\mathrm{a}}^{2}}. 
\label{peiwf}
\end{equation}
The right-hand side of the equation is quite close to the Lorenz number, 
$L_0 = (\pi^{2}/3)(k_{\mathrm{B}}/e)^{2}$. This implies  that the phonon
thermal conduction at high temperatures is comparable with the electronic
thermal conduction, if both of them are determined  by the electron-phonon
scattering only \cite{ziman}.  

\subsection{Superconductivity effects in thermal conductivity} 
\label{super}
In superconductors, as temperature crosses $T_{\mathrm{C}}$ from above, some 
number of electrons, $n_{\mathrm{s}}$, becomes superconducting, while the rest 
of them, in number $n_{\mathrm{n}}$, would remain as before in the normal 
state. The fraction of superconducting electrons [given by 
$x=n_{\mathrm{s}}/(n_{\mathrm{s}}+n_{\mathrm{n}})$] increases  continuously 
with decreasing temperature from $x=0$ at 
$T=T_{\mathrm{C}}$ to $x=1$ at $T=0$. The superconducting electrons are in 
a bound state (Cooper pairs), in which they are unable to transport entropy 
or interact with phonons. But the rest of the electrons (the fraction,
$1-x$) being normal, remain heat carriers and can interact with the phonons.
\par
Even in the frame of this simple picture (which is in the spirit of the 
two-fluid model of Gorter and Casimir) some quite definite predictions about 
the effects of superconductivity on the thermal conductivity can be made. 
Before doing this, let us denote the electronic thermal conductivity in the
superconducting state as $\kappa_{\mathrm{e}}^{\mathrm{s}}$, and that in 
the normal state (induced, for example, by high enough magnetic field) as 
$\kappa_{\mathrm{e}}^{\mathrm{n}}$.  The same kind of notations 
($\kappa_{\mathrm{p}}^{\mathrm{s}}$  and $\kappa_{\mathrm{p}}^{\mathrm{n}}$)
will be used in the following for the phonon part of the thermal conductivity 
or even for the total thermal conductivity ($\kappa^{\mathrm{s}}$  
and $\kappa^{\mathrm{n}}$). It is clear that the ratio 
$\kappa_{\mathrm{e}}^{\mathrm{s}}/\kappa_{\mathrm{e}}^{\mathrm{n}}$ must 
reduce  continuously  with decreasing temperature below $T_{\mathrm{C}}$ from 
the initial value (equal to unity) at $T=T_{\mathrm{C}}$ to values much less 
than unity at low enough temperature. All this takes place due to a decrease 
in the fraction of the normal electrons when going below $T_{\mathrm{C}}$. 
The same reason (that is a decrease in the number of the normal electrons 
below $T_{\mathrm{C}}$) can, however, induce an increase in the ratio, 
$\kappa_{\mathrm{p}}^{\mathrm{s}}/\kappa_{\mathrm{p}}^{\mathrm{n}}$, for 
the phonon contribution, causing it grow far above unity with decreasing 
temperature in the range not too far below $T_{\mathrm{C}}$. In this case a 
reduction in the number of normal electrons (which are the main phonon 
scatterers in metals for low temperatures) can cause a decrease in the 
phonon-electron relaxation rate, $1/\tau_{\mathrm{pe}}$, and, hence, an 
increase in the phonon thermal conductivity (it follows from Eq. (\ref{pei}) 
that $\kappa_{\mathrm{pe}}\propto 1/n_{\mathrm{e}}$).
\par
\begin{figure}[tbh]
\vspace{-10pt}
\centerline{\epsfig{file=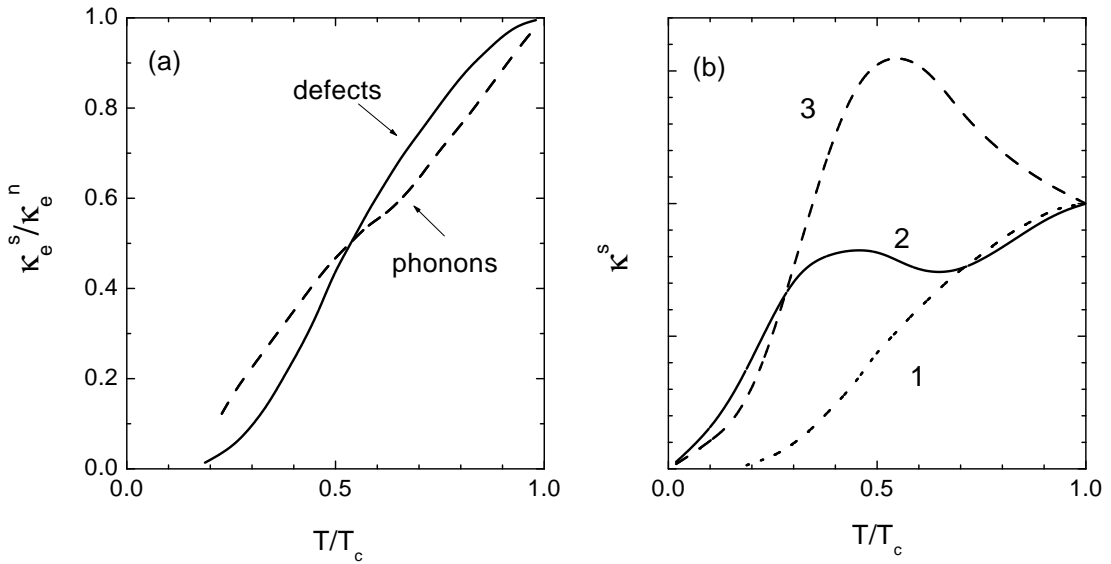,width=15cm}}
\caption{{\small {\sl Left panel (a) presents theoretical temperature 
dependences of the ratio of electronic thermal conductivities in 
superconducting and normal states, 
$\kappa_{\mathrm{e}}^{\mathrm{s}}/\kappa_{\mathrm{e}}^{\mathrm{n}}$, 
below $T_{\mathrm{C}}$ for the cases when the electrons are scattered 
predominantly by defects (solid line) or by phonons (dashed line) 
\cite{geilik,brt,gk,tewordt,gk2}. 
Right panel (b) presents schematic sketches of temperature dependences of the 
total thermal conductivity below $T_{\mathrm{C}}$ for different relationships 
between the electron and phonon contributions to the thermal conductivity. 
Expected behaviors for the cases when the electrons or phonons dominate the 
thermal conductivity are presented by curves 1 (doted line) and 3 (dashed 
line), respectively. Curve 2 (solid line) presents the intermediate case when 
both kinds of the heat carriers compete on equal terms in the thermal 
conductivity.}}} 
\label{figbrt}
\vspace{10pt}
\end{figure}

The basic theoretical results on the superconductivity effects in the thermal 
conductivity were obtained mainly in the frame of the BCS model 
\cite{geilik,brt,gk,tewordt,gk2,rick,ginsberg}. Consider them separately for 
the electron and phonon contributions to the thermal conductivity.  It turns 
out to be rather important for the electron heat transport whether the
critical temperature $T_{\mathrm{C}}$ falls  below or above the peak in the 
temperature dependence of $\kappa_{\mathrm{e}}(T)$. In the former case, 
the normal electrons (or as it is often said, quasiparticle excitations) are 
scattered predominantly by the lattice imperfections, while in the latter 
case, by phonons. For both cases rather cumbersome expressions for temperature
dependences of 
$\kappa_{\mathrm{e}}^{\mathrm{s}}/\kappa_{\mathrm{e}}^{\mathrm{n}}$ were 
derived which can be found in 
Refs. \cite{uher,geilik,brt,gk,tewordt,gk2,rick,ginsberg}. We shall restrict 
ourselves to consideration of the graphical representations of these 
dependences shown in Fig. \ref{figbrt}(a). When defects are the dominant 
electron scatterers, the function  
$\kappa_{\mathrm{e}}^{\mathrm{s}}/\kappa_{\mathrm{e}}^{\mathrm{n}} = f(T)$
has a zero slope at $T_{\mathrm{C}}$ [solid curve in Fig. \ref{figbrt}(a)]; 
whereas, in the case that electrons are scattered mainly by phonons [dashed 
curve in Fig. \ref{figbrt}(a)], the slope at $T_{\mathrm{C}}$ is rather large 
(about 1.62 according to Ref. \cite{tewordt}). Theories 
\cite{geilik,brt,gk,tewordt,gk2,rick,ginsberg}, including the so called BRT 
model \cite{brt,tewordt,rick}, agree well with experiment for the case of 
predominantly electron-imperfection scattering that is found for pure 
low-$T_{\mathrm{C}}$ superconductors with high electron density like Al, In, 
Sn and some others (see \cite{parrot,tewordt,ginsberg}).  For the case where 
elecron-phonon scattering dominates, the agreement between theory and 
experiment is not so convincing.
\par
It should be noted that for a comparison between the theories 
\cite{geilik,brt,gk,tewordt,gk2} and experimental data one must know exactly 
the ``normal'' behavior of thermal conductivity below $T_{\mathrm{C}}$ since 
a theoretical expression for     
$\kappa_{\mathrm{e}}^{\mathrm{s}}/\kappa_{\mathrm{e}}^{\mathrm{n}}=f(T)$ is 
used for the comparison. For a weak superconductor with low $T_{\mathrm{C}}$, 
the ``normal'' thermal conductivity can be measured below $T_{\mathrm{C}}$ 
under a magnetic field which is high enough to suppress superconductivity. 
This is not feasible, however, for superconductors (like cuprates) with high 
$T_{\mathrm{C}}$ and huge critical magnetic fields. Besides, when 
$T_{\mathrm{C}}$ is high enough (as in the cuprates) and falls into the 
temperature range where electron-phonon scattering dominates, the WF law fails 
 and does not give a good estimate of the electronic part of the thermal 
conductivity  from the total measured thermal conductivity (see 
Sec. \ref{subetc}). The comparison with BRT and other models in the  case of 
superconductors with high enough $T_{\mathrm{C}}$ poses, therefore, great 
difficulties.

\par
The phonon contribution to heat transport can be quite appreciable for some 
materials and be comparable or even higher than that of the electrons. In 
this case the thermal conductivity behavior below $T_{\mathrm{C}}$ can be
partially or totally determined by phonon transport.  Figure \ref{figbrt}(b) 
presents schematic sketches of possible temperature dependences of the total 
thermal conductivity, $\kappa^{\mathrm{s}}$,  below $T_{\mathrm{C}}$ for 
different relationships between the electron and phonon contributions to the 
thermal conductivity. A  behavior of $\kappa^{\mathrm{s}}(T)$ for a limiting 
case when the electron contribution dominates is presented by curve 1 (doted 
line). This type of behavior is expected for rather perfect superconducting 
metals with a high electron concentration.  Another limiting case, when 
phonons dominate in the thermal conductivity, is illustrated by curve 3 
(dashed line). 
In this case (as was already mentioned above), the phonon thermal conductivity 
increases below $T_{\mathrm{C}}$ with decreasing temperature due to the 
decrease in the number of normal electrons with a corresponding decrease in 
the phonon-electron relaxation rate. At low enough temperature the phonons 
begin to be scattered primarily by crystal boundaries, so that the phonon mean 
free path becomes temperature independent. After that, according to general 
relation (\ref{eq2}), $\kappa^{\mathrm{s}}(T)$ decreases to zero with 
decreasing temperature with the resulting formation of a maximum in 
$\kappa^{\mathrm{s}}(T)$ dependence [curve 3 in Fig. \ref{figbrt}(b)].  This 
type of the  $\kappa^{\mathrm{s}}(T)$-behavior is expected in disordered 
metals and alloys with high electron-imperfection scattering (that depresses 
the electron contribution to the thermal conductivity), and was seen in some 
disordered low-$T_{\mathrm{C}}$ superconductors \cite{gk2,mendel}. This 
behavior was found to be in good agreement with the theory of Geilikman and 
Kresin \cite{gk2}. The phonon thermal conductivity may be dominant also in 
rather perfect metals which have a low free electron density. This is 
characteristic of high-$T_{\mathrm{C}}$ cuprates where 
$\kappa^{\mathrm{s}}(T)$-behavior, similar to that of curve 3 in 
Fig. \ref{figbrt}(b), is frequently observed \cite{uher}.  
\par
The curves 1 and 3 in Fig. \ref{figbrt}(b) present the limiting cases when one 
type of heat carriers (electrons or phonons) dominates the thermal 
conductivity. It is apparent that, for many materials, the electron and phonon 
contributions will be of the same order. This will lead to dependences which 
are intermediate between the two limiting cases. Curve 2 (solid line) in 
Fig. \ref{figbrt}(b) 
presents one of the possible intermediate cases when both kinds of the heat 
carriers compete on equal terms in the thermal conductivity. Some examples of 
this kind of $\kappa^{\mathrm{s}}(T)$ behavior for low-$T_{\mathrm{C}}$ 
superconductors can be found in Ref. \cite{gk2}.

\section{Results and Discussion for Magnetic Superconductors}
\label{magnsuper}
\subsection{Rare-earth Nickel Borocarbides}
\label{boro}
In this section we consider the thermal conductivity data for rare-earth 
nickel borocarbides, RNi$_2$B$_2$C, with R = Y, Lu, Yb, Tm, Er, Ho, Dy, Tb  
or Gd (which will be referred below to as borocarbides). Although only those 
members with R = Tm, Er, Ho and Dy are magnetic superconductors, for the sake 
of completeness and to illustrate similarities for the members that exhibit
only one of these behaviors (only superconductivity, only magnetism) or 
even neither, all members will be discussed. Before going directly 
to this matter, it is appropriate first to briefly review the basic properties 
of these compounds. For extended reviews of experimental and  theoretical 
findings for borocarbides see Refs. \cite{hilscher,don,drec,muller}.

\subsubsection{\footnotesize{\bf General Properties}} 
\label{general}
The crystal structure of RNi$_2$B$_2$C is a body-centered tetragonal  
with space group $I4/mmm$ \cite{sieg}. It is a layered structure in which   
Ni$_2$B$_2$ layers are separated by R-C planes stacked along the $c$-axis. The 
Ni atoms are tetrahedrally coordinated to four B atoms. Despite the layered 
structure,  the electronic properties of borocarbides appear to be essentially 
three-dimensional according to electronic band structure calculations 
\cite{ebscal}, supported by the spectroscopic studies \cite{spectr}. These 
compounds are rather good metals  with a large electron density of states 
$N({\cal E}_{\mathrm{F}})$ at the Fermi level. The dominant contribution to  
$N({\cal E}_{\mathrm{F}})$ comes from Ni $3d$-states, but some contributions 
come from all other atoms as well \cite{ebscal}. Resistivity and 
magnetoresistance of RNi$_2$B$_2$C borocarbides (R = Y, Lu, Er, Ho) 
\cite{fisher} have not revealed appreciable anisotropy between measurements 
normal to the $ab$-plane (along the $c$-axis) and in-plane, consistent with 
the three-dimensional character of the electronic 
properties.    Small deviations from isotropic behavior were observed below 
150 K for the compounds containing magnetic ions (Er and Ho), which was 
attributed in Ref. \cite{fisher} to crystal-electric field effects. 
\par
Among the  RNi$_2$B$_2$C family, the compounds with R = Y, Lu, Yb are 
non-magnetic (the last is a heavy-fermion system), but all others 
(R = Gd, Tb, Dy, Ho, Er, and Tm) are magnetic. Their magnetic properties  are 
determined by the localized electrons in the $4f$-shell of R atoms. Long range 
magnetic order is due to the indirect RKKY interaction, mediated through the
conduction electrons. This gives  
rise to different types of antiferromagnetic (AFM) order of the $4f$-ions 
\cite{lynn}. Borocarbides with R = Tm, Er, Ho, Dy show coexistence of 
superconductivity and magnetic order. More information about superconducting 
and magnetic properties of these compounds can be found in Table~I. 
\begin{table}[tbh]
\caption[Transitions of the RNi$_2$B$_2$C bulk compounds]{{\sl Low temperature 
phase states of the RNi$_2$B$_2$C bulk compounds (taken from Refs. 
\cite{don,muller,lynn} unless otherwise referenced). R~=~Y or rare-earth 
element, T$_{\mathrm{C}}$~ is the superconducting transition temperature, 
T$_{\mathrm{M}}$ is the magnetic transition temperature, $T_{\mathrm{K}}$ is 
Kondo temperature. Column ``Magnetic Transition'' indicates magnetic states 
into which the compounds can be transformed (AFM~=~antiferromagnet, 
MAFM~=~modulated incommensurate antiferromagnet and WFM~=~weak ferromagnet), 
$\vec q$ is the wave vector of the modulated magnetization (in units of
reciprocal lattice parameters) and the symbol $\uparrow \parallel$ indicates 
magnetic  moment directions. Abbreviation HFS denotes heavy-fermion system.}}
\label{transitions}
\begin{center}
\begin{tabular}{|c|c|c|l|}
\hline
R	&$T_{\mathrm{C}}$ (K)	&$T_{\mathrm{M}}$ (K)	&Magnetic Transition\\ \hline
\hline
Y	&15.7		&0		&none\\ \hline
Lu	&16.6		&0		&none\\ \hline
Yb	&$<$0.05	&$<$0.023 \cite{bon}	&HFS,
T$_K$ = 10 K, $\gamma$ = 530 mJ/mol-K\\ \hline
Tm	&11		&1.5		&MAFM~~ $\vec q$ $\approx$ 0.094(a$^\ast$$\pm$b$^\ast$) $\uparrow \parallel <\!001\!>$\\ \hline
Er	&11		&6.8		&MAFM~~ $\vec q$ $\approx$ 0.55a$^\ast$ $\uparrow \parallel <\!010\!>$ \\
	&		&2.3		&WFM~~ $\uparrow \parallel <\!100\!>$~or~$ <\!110\!>$ \cite{canf}   \\ \hline
Ho	&8.5		&6.0		&MAFM~~ $\vec q$ $\approx$ 0.585a$^\ast$, 0.915c$^\ast$\\
	&		&5.5		&MAFM~~ $\vec q$ $\approx$ ?\\
	&		&5.2		&AFM~~ $\vec q$ = c$^\ast$ $\uparrow \parallel <\!110\!>$\\ \hline
Dy	&6.2		&10.3		&AFM~~ $\vec q$ = c$^\ast$ $\uparrow \parallel <\!110\!>$\\ \hline
Tb	&$<$0.3		&14		&MAFM~~ $\vec q$ $\approx$ 0.55a$^\ast$$\uparrow \parallel <\!100\!>$  \\
	&		&6-8		&WFM~~ $\uparrow \parallel <\!100\!>$~or~$<\!110\!>$ \cite{cho} \\   \hline
Gd	&$<$0.3		&20		&MAFM~~ $\vec q$ $\approx$ 0.55a$^\ast$ $\uparrow \parallel <\!010\!>$\\
	&		&13.6		&Tilted MAFM\\ \hline
\end{tabular} 
\end{center}
\end{table}
\par
A Cooper pair consists of two electrons with equal and opposite moment,
and with opposite spins (the total momentum and spin of the pair are zero). 
Any perturbation which acts with opposite signs (or with opposing force) on 
the two members of a Cooper pair can destroy this pair, producing a, so 
called, 
pair breaking effect. For example, an external or the internal (in magnetic 
materials) magnetic field exerts an orbital pair-breaking effect since the 
field acts with opposing force on the two electron momenta in the pair. If 
some ions in a metal system have a magnetic moment, ion spins will act with 
opposite sign on the electron spins in the pair (the magnetic pair-breaking). 
Generally speaking \cite{degen}, such types of perturbations lead to breaking 
of the time-reversal properties of the system and, hence, can cause a strong 
decrease in $T_{\mathrm{C}}$ or even total suppression of superconductivity.
\par 
In the case of rare-earth compounds, localized  $4f$-electrons should 
undoubtely exert a pair-breaking effect on the superconductivity. Since,
however, the  $4f$-electrons are strongly localized in deep inner 
$4f$-orbitals, their interaction with conducting electrons can be rather
weak, thus, permitting coexistence of superconductivity and the long range 
magnetic order. Such a situation is believed to take place for rare-earth 
rhodium boride compounds, Chevrel phases \cite{maple,bulai,fischer}, and  
borocarbides \cite{hilscher}. Coexistence of superconductivity and AFM order 
was justified theoretically rather long ago \cite{balten}. The coexistence is
possible if the AFM exchange field averages to zero within the superconducting 
coherence length, $\xi$. AFM order is, however, expected to have a profound 
effect on superconducting and transport properties, and this was found in the 
above-mentioned magnetic superconductors. 
\par
The theoretical and experimental studies generally indicate that the 
borocarbides are 
conventional electron-phonon mediated superconductors with $s$-wave symmetry 
for the order parameter \cite{hilscher,spectr}. On the other hand, some 
members of the borocarbide family manifest properties which are suggestive of 
unconventional or exotic superconductors \cite{drec,brandow,annett,maki}. In 
spite of this, the borocarbides are reasonably considered to be conventional 
superconductors. 
\par
Rare-earth borocarbides are rather good conductors, comparable to transition 
metals or their alloys. They have high electron density. According to 
Ref. \cite{mandal}, the value of $n_{\mathrm{e}}$ for Y-, Ho-, and Gd-based 
borocarbides are 2.63, 3.12 and $5\times 10^{22}$~cm$^{-3}$, respectively.
These values are only moderately less than those of simple superconducting 
metals, such as Al, Pb or Sn \cite{ashcroft}. The Fermi velocity values, 
$3.6\times 10^{7}$~cm/s (calculated for LuNi$_{2}$B$_{2}$C by Pickett and 
Singh \cite{ebscal}), and $4.2\times 10^{7}$~cm/s (found by de Haas-van Alphen 
studies in YNi$_{2}$B$_{2}$C \cite{winzer}) are, however,  clearly much less 
than the typical value $v_{\mathrm{F}} \approx 2\times 10^{8}$ cm/s for simple 
metals \cite{ashcroft}. Measurements of transport properties of borocarbide 
single crystals \cite{don,fisher,don2} have shown that the resistivity at room 
temperature, $\rho_{\mathrm{RT}}$, is typically in the range 35--70 
$\mu \Omega\, \mathrm{cm}$, and that the residual resistivity, $\rho_{0}$, is 
in the range 2--5 $\mu \Omega\, \mathrm{cm}$ ($T < 10-15$~K), so that the 
ratio $\rho_{\mathrm{RT}}/\rho_{0}$ is in the range 10--30. These conducting 
properties, though rather good, are not so good as those of the pure simple 
metals, for which $\rho_{0}$ can be as low as 0.01 $\mu \Omega\, \mathrm{cm}$ 
and $\rho_{\mathrm{RT}}/\rho_{0}$ can be as high as a thousand \cite{meaden}.  
It thus follows that the relative electron contribution to the total thermal 
conductivity in borocarbides should be much less than that in pure simple 
metals.  For this reason, some manifestations of the phonon contribution in 
the behavior of the total thermal conductivity of borocarbides is expected. In 
particurlar, due to the rather high residual resistivity, $\rho_{0}$, the 
phonon effects may show themselves below 
$T_{\mathrm{C}}$.  The high $\rho_{0}$-values even in single-crystals are 
possibly determined by some types of inhomogeneities and point defects. 
The known studies \cite{nagara2} suggest vacancies at the boron/carbon sites,
which have no appreciable influence on bulk superconducting properties. The 
vacancies can raise, however, $\rho_{0}$-values considerably, depressing
in this way the electronic thermal conductivity and enhancing the role of 
the phonon thermal conductivity in the low temperature range.  
\par
The temperature dependence of the resistivity in borocarbides deserves some  
attention since it may be important for consideration of their thermal 
conductivity. The experimental $\rho (T)$ curves are found to be 
approximately linear in the range  100--300~K (see $\rho(T)$ plot for 
LuNi$_2$B$_2$C in Fig. \ref{ybrho}) \cite{don,don2}. Below 50 K,
$\rho (T)$ is essentially non-linear and, in a rather narrow range 
$1.25\, T_{\mathrm{C}}<T<$(30--40)~K, can be approximated by 
$\rho(T)= \rho_{0} + aT^{\mathrm{p}}$, where ${\tt p}$-values for 
RNi$_2$B$_2$C compounds are in the range $2.0\alt \mathrm{p}\alt 2.6$ 
\cite{don,don2}.  
\par
Since the specific heat enters the general expression (\ref{eq2}) for the 
thermal conductivity, a brief mention should be made of this point. Most  
studies focussed on the specific heat in the neighborhood of $T_{\mathrm{C}}$ 
(above and below it) \cite{hilscher,carter,hong, movshe}. The lattice specific 
heat, $C_{\mathrm{p}}$, in fairly good metals begins to exceed the electronic 
contribution at a temperature which is a few percent of the Debye temperature 
\cite{ashcroft}, and the same must be true for the borocarbides as well. In 
the Debye model, $C_{\mathrm{p}}$ tends to a constant value when $T$ 
approaches $\Theta_{\mathrm{D}}$. It is essential, therefore, to know the 
$\Theta_{\mathrm{D}}$-values for the compounds considered. In the specific heat
measurements the following values of $\Theta_{\mathrm{D}}$ were obtained: 
345 K for R = Lu \cite{carter}, $\approx 540$~K \cite{hong} or $\approx 490$~K 
\cite{movshe} for R = Y, and 320 K for R=Tm \cite{movshe}. It is evident that 
$\Theta_{\mathrm{D}}$-value for other borocarbides fall within the outlined 
temperature range. Since most of the thermal conductivity measurements 
considered below were made in the range 4.2--300~K, it is important to know 
how $C_{\mathrm{p}}$ behaves in this range, especially in the high-temperature 
range 100--300 K. Judging from the above-indicated 
$\Theta_{\mathrm{D}}$-values, a fairly large rise in  $C_{\mathrm{p}}$ with 
increasing temperature should take place in the high-temperature range. This 
was found in LuNi$_2$B$_2$C \cite{michor}, where $C_{\mathrm{p}}$ has 
nearly doubled in the range 150--300 K. Although no other 
specific-heat studies were done (to our knowledge) in the high-temperature 
range, the same kind of the $C_{\mathrm{p}}(T)$ behavior must be expected for 
the other borocarbides as well.  

\subsubsection{\footnotesize{\bf Thermal Conductivity}}
\label{borotherm} 
The thermal conductivity studies of RNi$_2$B$_2$C compounds are few in number 
\cite{sera,cao,boak,brian3,boak2,brian4,izawa}. Some of them 
\cite{sera,cao,boak,boak2,izawa} have been made only at very low temperature. 
Below we consider results on single-crystal borocarbides with R = Y, Lu, Yb, 
Tm, Er, Ho, Dy, Tb, and Gd. Certain of the results have been partially 
presented in Refs. \cite{brian2,brian3,brian4}. The samples were provided by 
P. C. Canfield and his research group at Iowa State University. 

\newpage
\vspace{25pt}
{\bf {\footnotesize a) Borocarbide magnetic superconductors}}
\vspace{15pt}
\par
\begin{figure}[tbh]
\vspace{-30pt}
\centerline{\epsfig{file=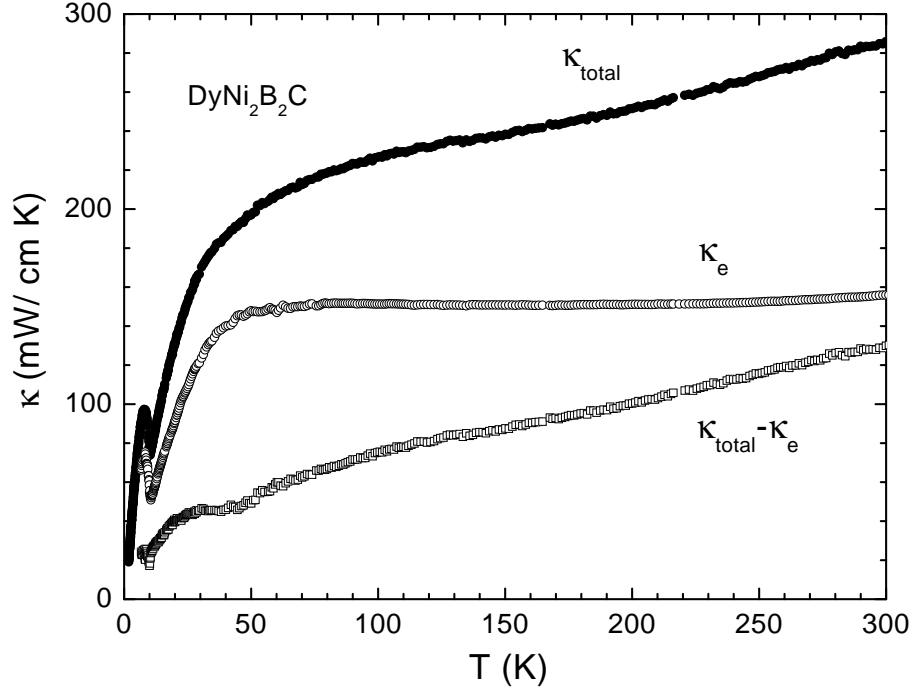,width=12cm}}
\caption{{\small {\sl Temperature dependence of thermal conductivity
($\kappa_{\mathrm{total}}$) for {\rm DyNi$_2$B$_2$C}. The meaning of
 $\kappa_{\mathrm{e}}$ and $\kappa_{\mathrm{total}}$--$\kappa_{\mathrm{e}}$ is 
explained in the main text of the article. }}} 
\vspace{10pt}
\label{dytot}
\end{figure}
\begin{figure}[tbh]
\vspace{-20pt}
\centerline{\epsfig{file=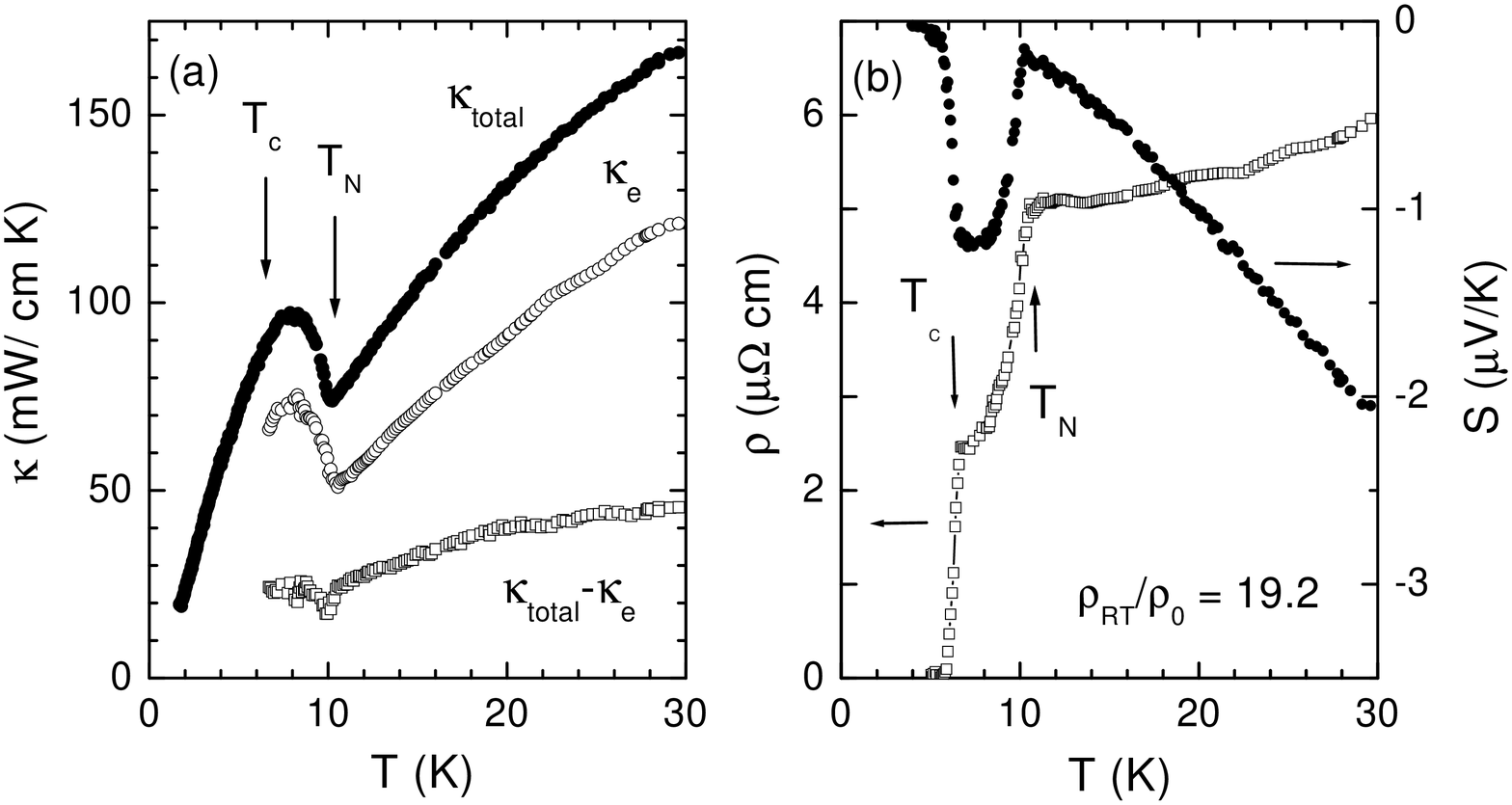,width=14cm}}
\caption{{\small {\sl $\kappa(T)$ {\rm (a)}, $\rho(T)$ and $S(T)$ 
{\rm (b)} for  {\rm DyNi$_2$B$_2$C} at low temperature.}}} 
\label{dydiff}
\end{figure}
\begin{figure}[tbh]
\vspace{-30pt}
\centerline{\epsfig{file=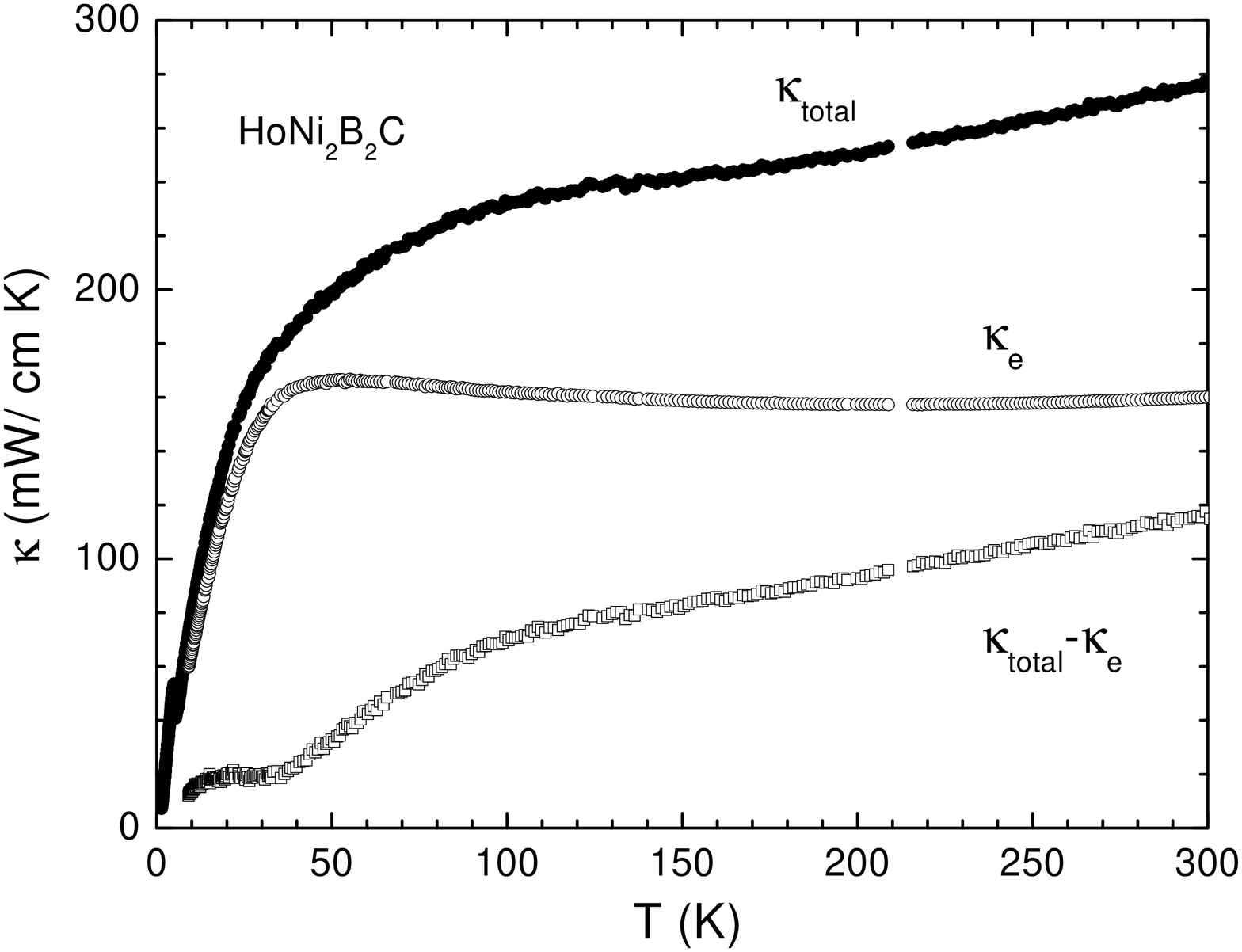,width=12cm}}
\caption{{\small {\sl Temperature dependence of thermal conductivity
($\kappa_{\mathrm{total}}$) for {\rm HoNi$_2$B$_2$C}. The meaning of
 $\kappa_{\mathrm{e}}$ and $\kappa_{\mathrm{total}}$--$\kappa_{\mathrm{e}}$ is 
explained in the main text of the article. }}} 
\vspace{10pt}
\label{hotot}
\end{figure}
\begin{figure}[tbh]
\vspace{-20pt}
\centerline{\epsfig{file=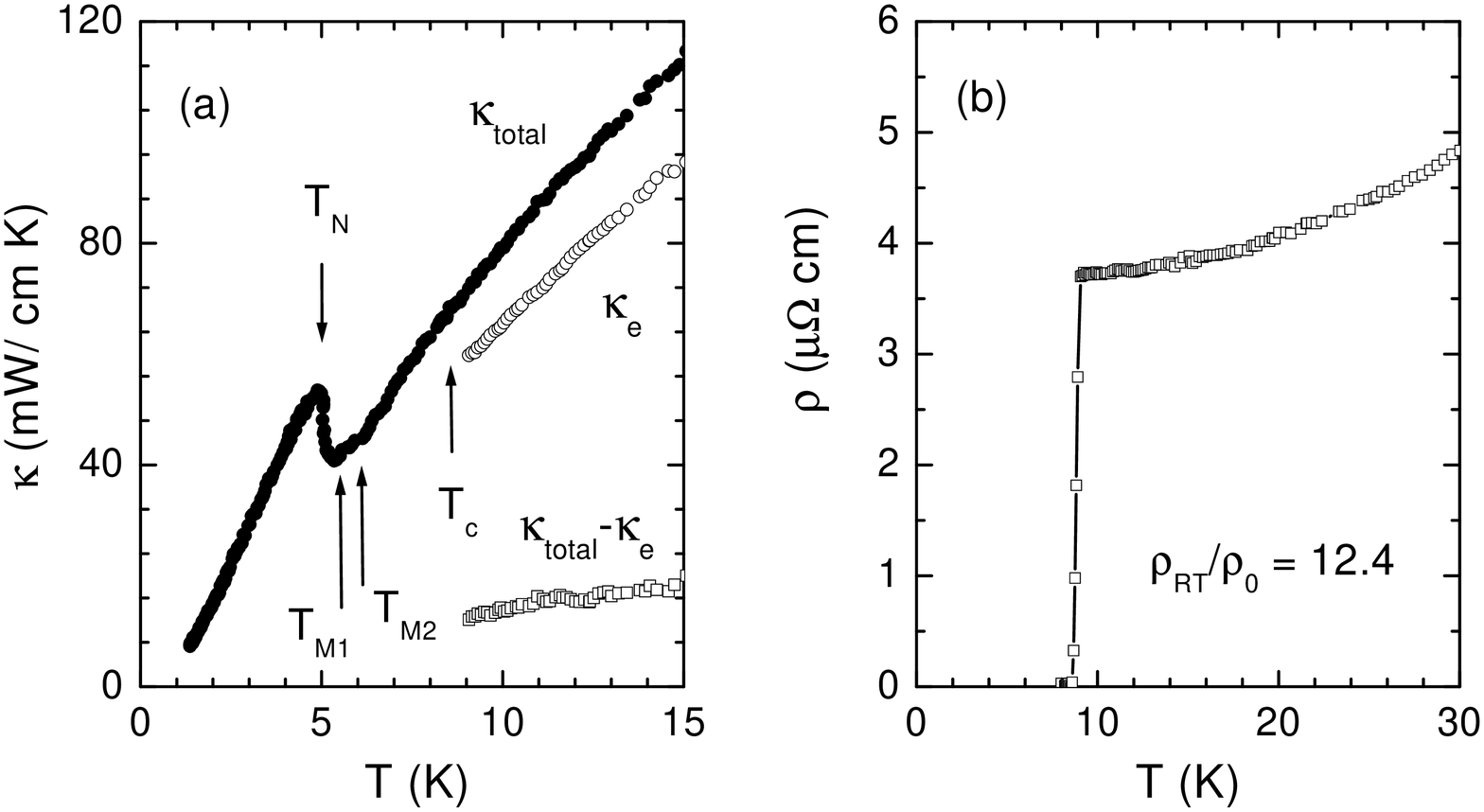,width=14cm}}
\caption{{\small {\sl $\kappa(T)$ {\rm (a)} and $\rho(T)$ {\rm (b)} 
for  {\rm HoNi$_2$B$_2$C} at low temperature.}}} 
\vspace{10pt}
\label{hodiff}
\end{figure}
\begin{figure}[tbh]
\vspace{-30pt}
\centerline{\epsfig{file=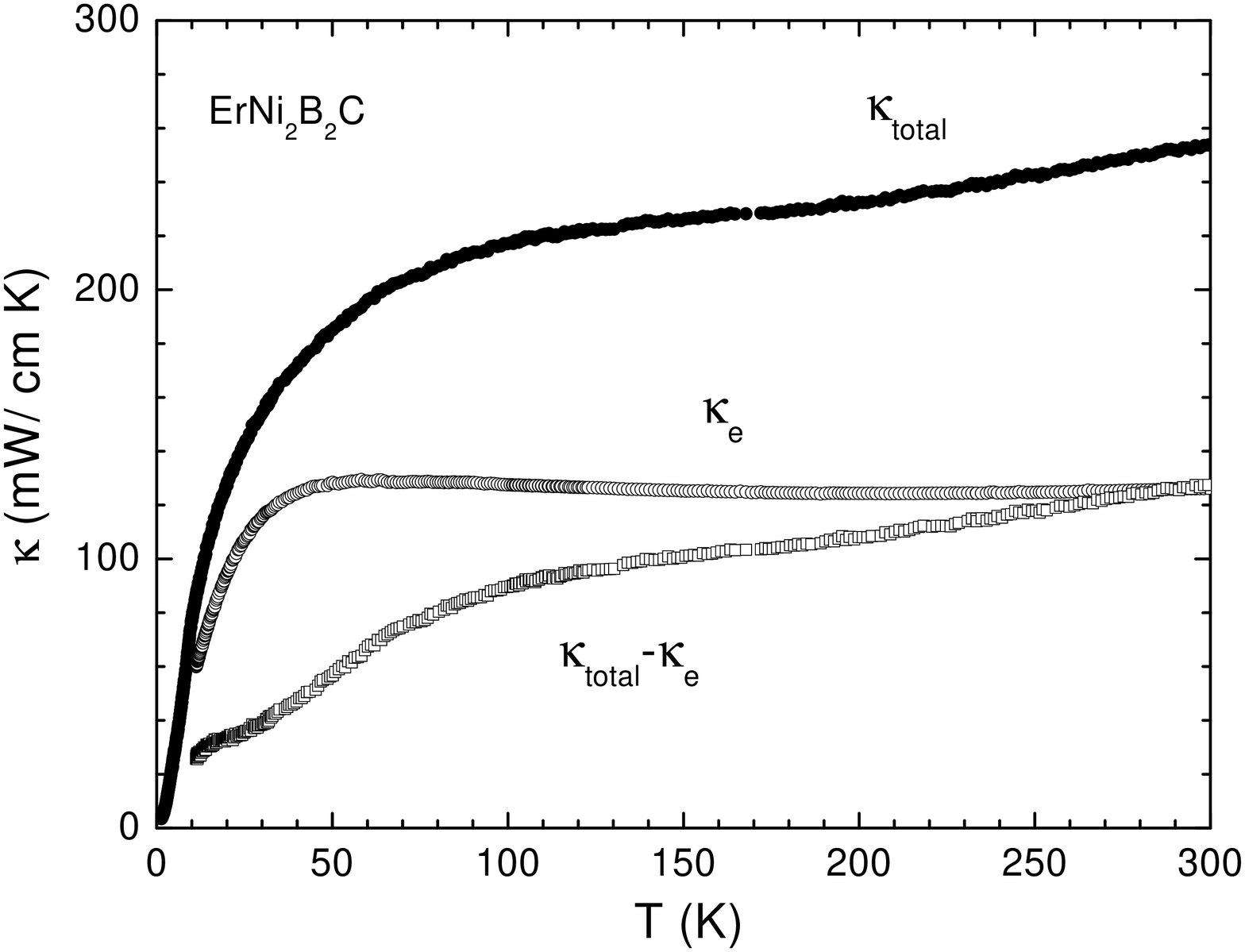,width=12cm}}
\caption{{\small {\sl Temperature dependence of thermal conductivity
($\kappa_{\mathrm{total}}$) for {\rm ErNi$_2$B$_2$C}. The meaning of
 $\kappa_{\mathrm{e}}$ and $\kappa_{\mathrm{total}}$--$\kappa_{\mathrm{e}}$ is 
explained in the main text of the article. }}} 
\vspace{10pt}
\label{ertot}
\end{figure}
\begin{figure}[tbh]
\vspace{-30pt}
\centerline{\epsfig{file=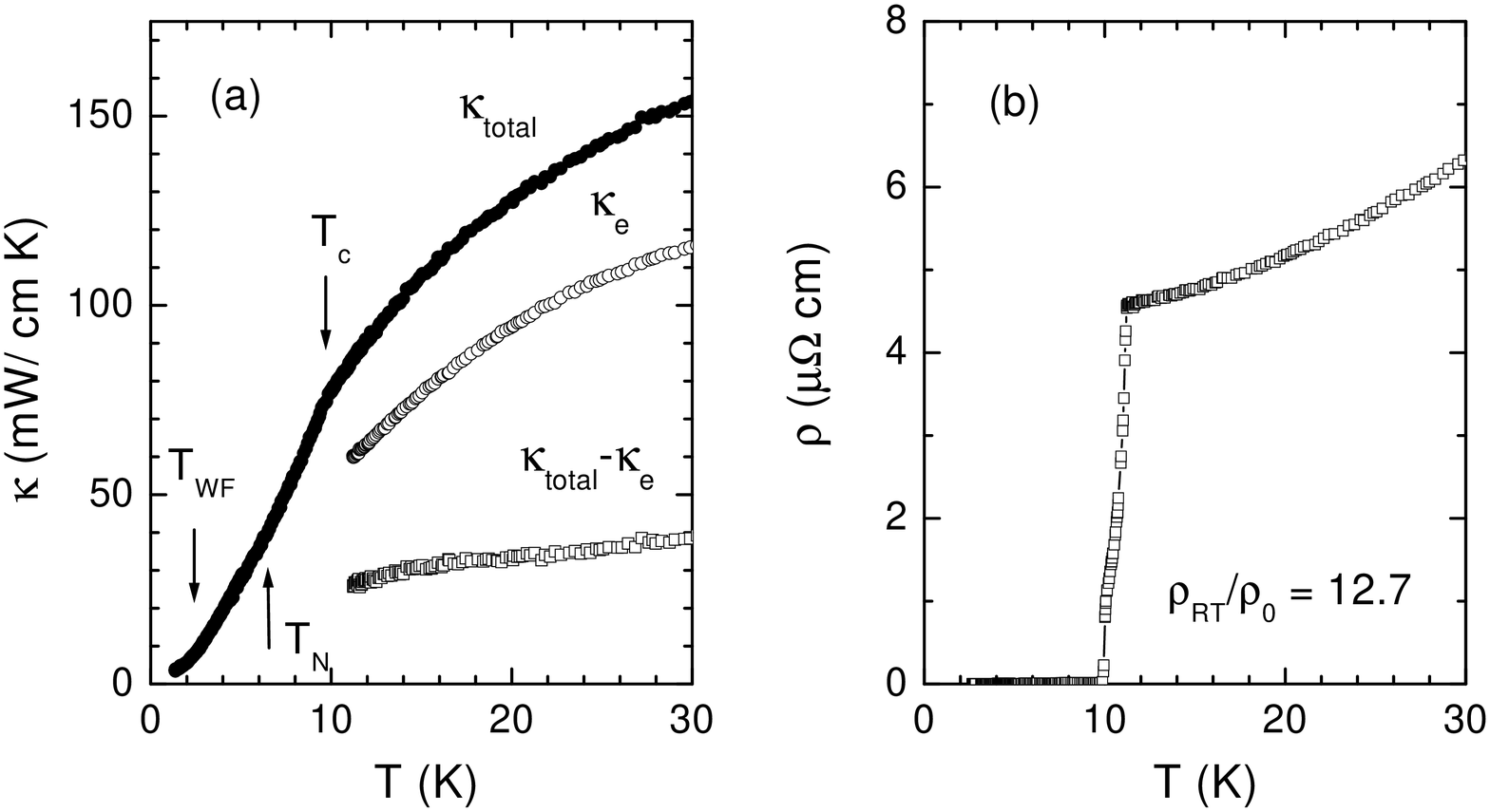,width=14cm}}
\caption{{\small {\sl $\kappa(T)$ {\rm (a)} and $\rho(T)$ {\rm (b)} 
for  {\rm ErNi$_2$B$_2$C} at low temperature.}}} 
\label{erdiff}
\end{figure}
\begin{figure}[tbh]
\vspace{-30pt}
\centerline{\epsfig{file=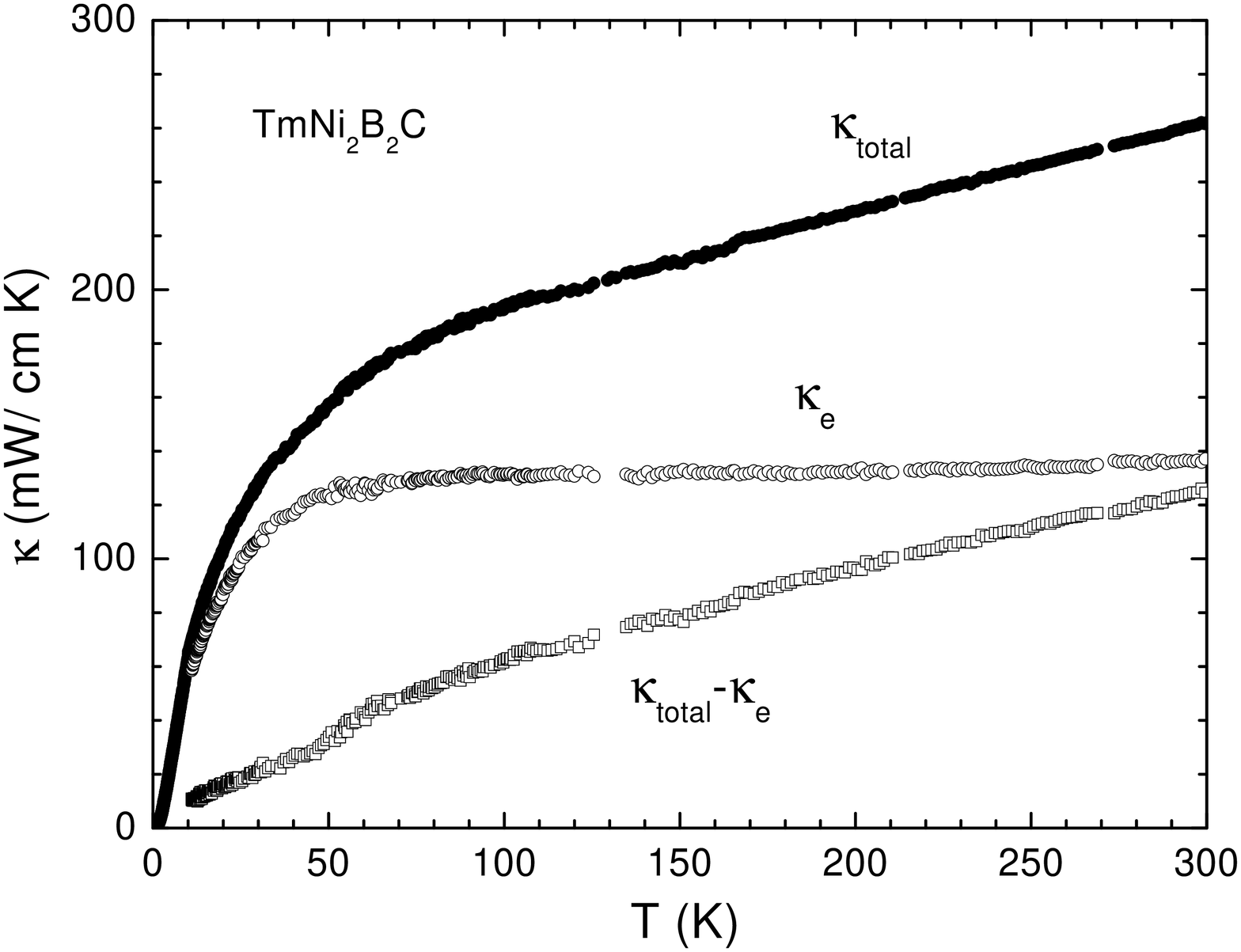,width=12cm}}
\caption{{\small {\sl Temperature dependence of thermal conductivity
($\kappa_{\mathrm{total}}$) for {\rm TmNi$_2$B$_2$C}. The meaning of
 $\kappa_{\mathrm{e}}$ and $\kappa_{\mathrm{total}}$--$\kappa_{\mathrm{e}}$ is 
explained in the main text of the article. }}} 
\vspace{10pt}
\label{tmtot}
\end{figure}
\begin{figure}[tbh]
\vspace{-30pt}
\centerline{\epsfig{file=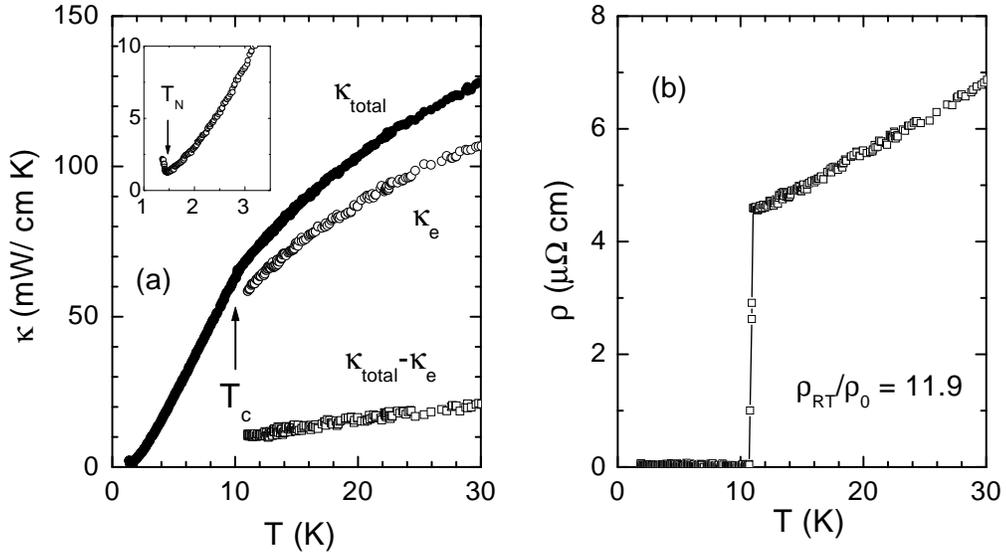,width=14cm}}
\caption{{\small {\sl $\kappa(T)$ {\rm (a)} and $\rho(T)$ {\rm (b)} 
for  {\rm TmNi$_2$B$_2$C} at low temperature.}}} 
\label{tmdiff}
\end{figure}

Let us start with the magnetic 
superconductors (R= Dy, Ho, Er, Tm). The temperature behavior of their thermal 
conductivity is presented in Figs. \ref{dytot}--\ref{tmdiff}. Half of the 
figures present expanded views of $\kappa (T)$ and $\rho (T)$  (and even the 
absolute thermopower, $S(T)$, in Fig. \ref{dydiff}) dependences at low 
temperature. This allows one to follow closely the changes in these properties 
at phase transitions.  On the panels with $\rho(T)$ curves, the values of 
$\rho_{\mathrm{RT}}/\rho_{0}$ are indicated for characterization of the sample 
quality.\footnote{$\rho_{0}$ here is the value at the temperature
 $T_{\mathrm{C}}^{+}$ at the onset of the resistive superconducting 
transition, not necessarily the residual resistivity. For R = Tm, Er, Ho it 
includes the contribution from spin disorder scattering.} 
On all figures, $\kappa_{\mathrm{total}}$ stands for the 
measured thermal conductivity, $\kappa_{\mathrm{e}}$ is the WF estimate for 
the electronic contribution (based on the measured $\rho$ for these samples), 
and  the remainder, $\kappa_{\mathrm{total}}\! -\! \kappa_{\mathrm{e}}$, can 
be thought as being the phonon contribution. 
\par
The {\bf DyNi$_2$B$_2$C} compound transforms from the paramagnetic (PM) into 
AFM state at the N\'{e}el temperature ($T_{\mathrm{N}}\approx 10.3$~K), which 
is considerably higher than $T_{\mathrm{C}}\approx 6.2$~K (see Table~I and 
Refs. \cite{lynn,dy1}), so that this compound goes to the superconducting 
state already being in the AFM state. The PM-AFM transition is of first order 
and  characterized by very distinct and intensive peak in the temperature 
dependence of the specific heat \cite{dy1}. It can be seen from Fig. 
\ref{dydiff} that the transition results in a reduction by nearly half in the 
resistivity, and in an appreciable increase in the thermal conductivity and 
thermopower (in the latter case, the increase is just enormous). The main 
reason for these dramatic changes is a decrease in the rate of electron 
scattering by the spin disorder in response to the PM-AFM transition, thus 
producing the pronounced structure in $\kappa(T)$, $\rho(T)$ and $S(T)$ shown 
in Fig. \ref{dydiff}. It is known that spin disorder can give a considerable 
contribution to the resistivity above $T_{\mathrm{N}}$ (see the discussion in 
Ref. \cite{don}). Actually, a  relevant spin-disorder term, 
$\rho_{\mathrm{sd}}$ (with the electron relaxation rate, 
$\tau_{\mathrm{sd}}^{-1}$) must be added into Matthiessen's relation 
(\ref{eeres}) for $T>T_{\mathrm{N}}$. This term can be reasonable above 
$T_{\mathrm{N}}$, but, for ideal spin alignment below $T_{\mathrm{N}}$, it 
may be thought to be equal to zero. The same reasoning is applicable to 
the thermal conductivity with the resulting addition of the term 
$\tau_{\mathrm{sd}}^{-1}$ to the relation (\ref{etres}). The relaxation rate, 
$\tau_{\mathrm{sd}}^{-1}$, is generally considered to be temperature 
independent above $T_{\mathrm{N}}$ \cite{gratz}. 
\par
It is evident from Fig. \ref{dydiff} that the main change in the
$\kappa_{\mathrm{total}}(T)$ at the PM-AFM transition can be attributed to 
that of the electronic contribution, $\kappa_{\mathrm{e}}$. Nevertheless, a 
small dip can be distinguished in the temperature curve of the phonon 
contribution, 
\newpage

\newpage

\newpage

\newpage

\newpage
\noindent 
$\kappa_{\mathrm{total}}\! -\! \kappa_{\mathrm{e}}$, in the
vicinity of $T_{\mathrm{N}}$, as well. The phonon thermal conductivity can be 
responsive to a magnetic transition when this is accompanied by strong enough 
magnetoelastic (magnetostriction) effects.
\par 
There is no easily discernable change in $\kappa_{\mathrm{total}}(T)$ at 
$T_{\mathrm{C}}$ (Fig. \ref{dydiff}). It is possible that rather 
small changes in slope of $\kappa(T)$ at $T_{\mathrm{C}}$ is somewhat hidden
on the background of the rapid variation in $\kappa(T)$ due to the PM-AFM 
transition. This variation can be appreciable even well below 
$T_{\mathrm{N}}$ down to $T_{\mathrm{C}}$. This is supported by the point that 
fairly large changes of $\rho(T)$ and $S(T)$ due to this transition occur in 
the temperature range between $T_{\mathrm{N}}$ and $T_{\mathrm{C}}$ 
[Fig. \ref{dydiff}(b)]. No sharp change should be expected either if the 
electron contribution, $\kappa_{\mathrm{e}}$, dominates, and the electrons are 
scattered predominantly by defects. No abrupt change in slope of 
$\kappa_{\mathrm{e}}(T)$ at $T_{\mathrm{C}}$ occurs in that case [see 
Fig. \ref{figbrt}(a) and discussion in Sec. \ref{super}]. 
\par
The {\bf HoNi$_2$B$_2$C} compound has three magnetic transitions below 
$T_{\mathrm{C}}\approx8.5$~K (see Table~I and Refs. \cite{lynn,ho1,ho2}), so
that a coexistence of bulk superconductivity and local long-range magnetic 
order takes place. The $\rho(T)$ shows a sharp superconducting transition 
[Fig.~\ref{hodiff}(b)], but $\kappa_{\mathrm{total}}(T)$ does not show a
noticeable change in behavior at $T_{\mathrm{C}}$ [Fig.~\ref{hodiff}(a)] (the
same is found in Ref. \cite{sera}). This may be related to defects as the 
dominant electron scattering mechanism as discussed for the case of 
DyNi$_2$B$_2$C (see above), but it is also may be connected with an anomalous 
superconducting state in the temperature range between 
$T_{\mathrm{C}} \approx 8.5$~K and 6.5 K, which was  indicated by 
point-contact tunneling measurements \cite{yanson}. Of these magnetic 
transitions, only the transition to the simple AFM state at 
$T_{\mathrm{N}}\approx 5.2$~K produces an appreciable feature  in the 
$\kappa_{\mathrm{total}}(T)$ curve [Fig.~\ref{hodiff}(a)]. The other two 
[their transition temperatures are denoted by $T_{\mathrm{M1}}$ and 
$T_{\mathrm{M2}}$ in Fig.~\ref{hodiff}(a)] have a minimal effect on the 
thermal conductivity. A change in slope at $T_{\mathrm{M2}}$ is clearly 
visible, but any feature at $T_{\mathrm{M1}}$ is within the noise. It is 
interesting that the specific heat ($C$) behaves somewhat in a similar manner 
at these transitions \cite{ho1,ho2}. Whereas, a sharp and intensive peak in 
$C(T)$ is found at $T=T_{\mathrm{N}}$ \cite{ho1,ho2},  there are only 
weak discernible shoulders in the  $C(T)$ dependences at the temperatures 
$T_{\mathrm{M1}}$ and $T_{\mathrm{M2}}$ \cite{hilscher,ho1}. The reason is 
that AFM transition at $T_{\mathrm{N}}$ is of first order \cite{ho2}; whereas,
the other two are of second one. It is clear that first-order magnetic 
transitions, characterized by discontinuous changes in the magnetic entropy 
(see Refs. \cite{dy1,ho1} for borocarbides) and in the compound density, 
should lead to more significant change in the thermal conductivity of the 
borocarbides than the second-order ones (characterized by discontinuous
changes in the second derivatives of the Gibbs free energy, such as the 
volume thermal expansivity or the specific heat).   
\par
The {\bf ErNi$_2$B$_2$C} borocarbide transforms with decreasing temperature 
first into the superconducting state at $T_{\mathrm{C}}\approx 11$~K, then 
into a MAFM state at $T_{\mathrm{N}}\approx 6.8$~K (first order transition), 
and finally into a WFM state at $T_{\mathrm{WF}}\approx 2.3$~K (second order 
transition) (see Table I and Refs. \cite{lynn,canf,er}). For this compound a 
rather distinct change in the slope of $\kappa_{\mathrm{total}}(T)$ at 
$T_{\mathrm{C}}$~ can be seen [Fig. \ref{erdiff}(a)], consistent with 
measurements on polycrystalline ErNi$_2$B$_2$C sample by Cao {\it et al.} 
\cite{cao}. But no features can be distinquished in the 
$\kappa_{\mathrm{total}}(T)$ curve at $T_{\mathrm{N}}$ and $T_{\mathrm{WF}}$.  
\par
The {\bf TmNi$_2$B$_2$C} compound is superconducting below 
$T_{\mathrm{C}}\approx 11$~K, and undergoes a MAFM transition (of first order) 
below $T_{\mathrm{N}} \approx 1.5$~K (see Table~I and Refs. 
\cite{lynn,movshe,tm}). A rather clear kink is seen in the 
$\kappa_{\mathrm{total}}(T)$ curve at $T_{\mathrm{C}}$, which corresponds in 
Fig. \ref{tmdiff}(a) to the temperature of the superconducting transition from 
the $\rho(T)$ data. 
A sharp increase in the thermal conductivity is readily seen at about 1.4 K, 
that is slightly below $T_{\mathrm{N}}$ [see insert in Fig. \ref{tmdiff}(a)]. 
It is particurlarly remarkable that $\kappa_{\mathrm{total}}$ doubles in a 
very short temperature range. It is clear that this increase is determined by 
the transition to the AFM state. $T_{\mathrm{N}}$ is, however, much less than 
$T_{\mathrm{C}}$ ($T_{\mathrm{N}}/T_{\mathrm{C}}\approx 0.136$) and, 
therefore, the fraction of the  normal electrons (which remain heat carriers
below $T_{\mathrm{C}}$, see Sec. \ref{super}) must be very small below 
$T_{\mathrm{N}}$. For this reason, the explanation given above in the case of 
DyNi$_2$B$_2$C (an abrupt decrease in the rate of electron scattering by the 
spin disorder below $T_{\mathrm{N}}$), is inapplicable for  TmNi$_2$B$_2$C. 
Really, the theoretical calculations [see Refs. \cite{brt,tewordt} and 
Fig. \ref{figbrt}(a)] show that $\kappa_{\mathrm{e}}$ is negligibly small for 
$T_{\mathrm{N}}/T_{\mathrm{C}} < 0.2$.
\par
The transition of TmNi$_2$B$_2$C into the AFM state does not destroy its 
global superconductivity in the sense that the resistivity remains zero 
\cite{movshe}. But it could suppress the superconducting state in such way 
(causing the effective $T_{\mathrm{C}}$ to become less than 11~K) that the 
fraction of the normal electrons would increase rather abruptly below 
$T_{\mathrm{N}}$. Together with disappearing of the spin-disorder electron 
scattering below $T_{\mathrm{N}}$, this could lead to an appreciable increase 
in $\kappa_{\mathrm{e}}$. A considerable decrease in the upper critical field 
$H_{\mathrm{c2}}$ (parallel to the $ab$ plane) found in TmNi$_2$B$_2$C below
3 K \cite{don,don3} gives some support to this suggestion. According to 
Ref. \cite{tm}, the Tm moments order below 1.5 K with an incommensurate 
magnetic structure consisting of Tm moments, aligned ferromagnetically along 
the $c$ axis in the (110) planes, and the magnitude of the moments modulated 
sinusoidally along the diagonal of the $ab$ plane. This magnetic structure 
should, however, support magnons. The formation of the magnons, which are 
heat carriers and whose number is especially large in the temperature range 
just slightly below $T_{\mathrm{N}}$, also could explain the sharp rise in 
$\kappa_{\mathrm{total}}$ below $T_{\mathrm{N}}$. It is hard to judge at the 
moment, however, the possible relative contributions, if any, of these two 
inferred mechanisms without some additional experimental studies. 
\par
Let us survey now the general features of the thermal conductivity behavior 
in the borocarbide magnetic superconductors (R = Dy, Ho, Er, Tm) in the 
temperature range above both, $T_{\mathrm{C}}$ and $T_{M}$ (Figs. \ref{dytot}, 
\ref{hotot}, \ref{ertot}, and \ref{tmtot}). It can said that the electron 
contribution to the thermal conductivity, $\kappa_{\mathrm{e}}$, clearly 
dominates over the phonon one, 
$\kappa_{\mathrm{total}}$-$\kappa_{\mathrm{e}}$, only at low temperature 
($< 50$~K), but with increasing temperature the relative phonon contribution 
grows significantly, so that at room temperature these contributions are quite 
comparable (if not nearly equal).  The $\kappa_{\mathrm{e}}(T)$ is 
approximately constant for $T>100$~K, as expected according to Eq.~(\ref{etc}) 
for electronic thermal conductivity and the WF law (\ref{wf}) in the case of 
linear $\rho(T)$ dependence (which is approximately obeyed in this temperature 
range for all of the borocarbides mentioned). But there is no peak in the 
$\kappa_{e}(T)$ dependence which is expected below 0.1~$\Theta_{\mathrm{D}}$ 
for good enough metals (see Sec. \ref{subetc}). The reason is that the 
borocarbides are maybe not good enough metals to the extent that they should 
show this peak. Really, the highest thermal conductivity magnitude of 
borocarbides is generally below 0.3 W/cmK (see Figs. \ref{dytot}, \ref{hotot}, 
\ref{ertot}, \ref{tmtot} above and figures for other samples below). At the 
same time, the silver foil presented in Fig.~\ref{ag} has a peak of about 
14 W/cmK at $T_{\mathrm{max}}\approx 16$~K in the $\kappa(T)$ curve. The 
silver samples with a highest degree of perfection have a peak value as high 
as 200~W/cmK with $T_{\mathrm{max}}\approx 7$~K, but no peak at all can be 
found in the $\kappa(T)$ dependences of the fairly imperfect silver samples 
with thermal conductivity below 2--3 W/cmK \cite{childs}. An identical picture 
of the lattice-disorder influence was found for copper samples \cite{white}. 
The thermal conductivity of the borocarbides is expected to be far (at least, 
one order of magnitude) less than that of the noble metals, since these latter 
have higher electron density and their electrons have a far greater Fermi 
velocity.  The thermal conductivity 
of the transition metals is also far (one or even two order of magnitude) less 
than that of the noble metals. For these metals the peak in $\kappa(T)$ curves 
is not so sharp and is usually found only in the most perfect samples with 
$\kappa$ above 0.4--1.0 W/cmK \cite{childs}). In some transition metals with 
very low thermal conductivity (Mn and V) the peak was not found at all 
\cite{childs}. It can be suggested, therefore, that the absence of a peak for
the borocarbide $\kappa(T)$ curves is determined mainly by lattice defects.
\par
Consider now the temperature behavior of the phonon part, 
$\kappa_{\mathrm{total}}$-$\kappa_{\mathrm{e}}=\kappa_{\mathrm{p}}$, of the 
thermal conductivity, presented in Figs. \ref{dytot}, \ref{hotot}, 
\ref{ertot}, and \ref{tmtot} by the 
$\kappa_{\mathrm{total}}$-$\kappa_{\mathrm{e}}$ curves. According to the
general relation (\ref{eq7}), the $\kappa_{\mathrm{p}}$ is determined by 
phonon-phonon (PP), phonon-imperfection (PI) and phonon-electron (PE) 
interactions. The PP interaction should result in formation of a peak in the 
$\kappa_{\mathrm{P}}(T)$ dependence at 
$T_{\mathrm{max}} < 0.1$~$\Theta_{\mathrm{D}}$, which, however, can be 
depressed or totaly smeared by strong enough lattice disorder. The shoulders, 
clearly seen in Figs. \ref{dytot}, \ref{hotot}, and \ref{tmtot} for R = Dy, 
Ho, and Tm in the range 35--45 K, could be indicators of the peak in the case 
of rather strong lattice disorder. As temperature increases above 
$T_{\mathrm{max}}$, the PP interaction can only induce a decrease in 
$\kappa_{\mathrm{p}}(T)$  (see Sec. \ref{phonon}). This is inconsistent with 
the temperature behavior of the 
$\kappa_{\mathrm{total}}$-$\kappa_{\mathrm{e}}$ curves (Figs. \ref{dytot}, 
\ref{hotot}, \ref{ertot}, and \ref{tmtot}), which suggests that 
$\kappa_{\mathrm{p}}$ rises steadily with temperature increasing up to room 
temperature. This rise can be determined by other (PI and PE) mechanisms of 
phonon scattering. If the PI interaction is taken into account, the point 
defects can be assumed as the most important imperfections in the 
single-crystal borocarbides \cite{nagara2}. In the intermediate temperature 
range below $\Theta_{\mathrm{D}}$, the phonon wavelength decreases with 
increasing temperature approaching an interatomic distance. In this case the 
temperature dependence of the PI relaxation rate is expected to be rather 
weak. On the other hand, the specific heat, $C_{\mathrm{p}}$, of borocarbides 
rises significantly in this range (see discussion in Sec. \ref{general}). This 
can determine (or, at least, contribute to) the rise in $\kappa_{\mathrm{p}}$ 
with increasing temperature. Another source of this increase in 
$\kappa_{\mathrm{p}}$  
can be found in the PE interaction (see Sec. \ref{perel}). According to 
Eq.~(\ref{pei}), the corresponding term, $\kappa_{\mathrm{pe}}$, grows with
increasing temperature for all temperatures below $\Theta_{\mathrm{D}}$, and 
this growth can be profound in the intermediate temperature range. It is 
important as well that in this range the term $\kappa_{\mathrm{pe}}$ can be 
comparable with the electronic thermal conduction (see discussion in Sec. 
\ref{perel}). In this way the both, the PI interaction and the PE one, can 
contribute to the growth of the thermal conductivity of the borocarbides with 
increasing temperature.  

\vspace{25pt}
{\bf {\footnotesize b) Purely superconducting borocarbides}}
\vspace{15pt}

\par
Up to this point the properties  of magnetic superconductors 
(R = Dy, Ho, Er, Tm) has been considered. It is interesting to compare their 
behavior with that of the superconducting borocarbides without magnetic 
order, {\bf YNi$_2$B$_2$C} and {\bf LuNi$_2$B$_2$C}, with $T_{\mathrm{C}}$'s 
15.6 and 16.6 K, respectively. Both of them are believed to be clean-limit 
type II superconductors \cite{drec,muller,kim}. Neither Y nor Lu are magnetic
and neither YNi$_2$B$_2$C nor LuNi$_2$B$_2$C exhibit any magnetic order 
\cite{lynn}. No AFM correlations or local magnetic moments of nickel are 
present in YNi$_2$B$_2$C \cite{suh}.  Inelastic neutron scattering 
measurements show soft phonon modes below $T_{\mathrm{C}}$ for YNi$_2$B$_2$C 
\cite{kawano} and  LuNi$_2$B$_2$C \cite {derven}. A model phonon spectra was
presented in Ref. \cite{hilscher} based on specific heat measurements 
\cite{michor}. A very pronounced low temperature softening of the low energy 
optical phonons and the transverse acoustic modes was found. The thermal 
conductivity of these compounds was studied at low temperature in Refs. 
\cite{sera} (YNi$_2$B$_2$C) and \cite{boak,boak2} (LuNi$_2$B$_2$C).  
  
\begin{figure}[tbh]
\centerline{\epsfig{file=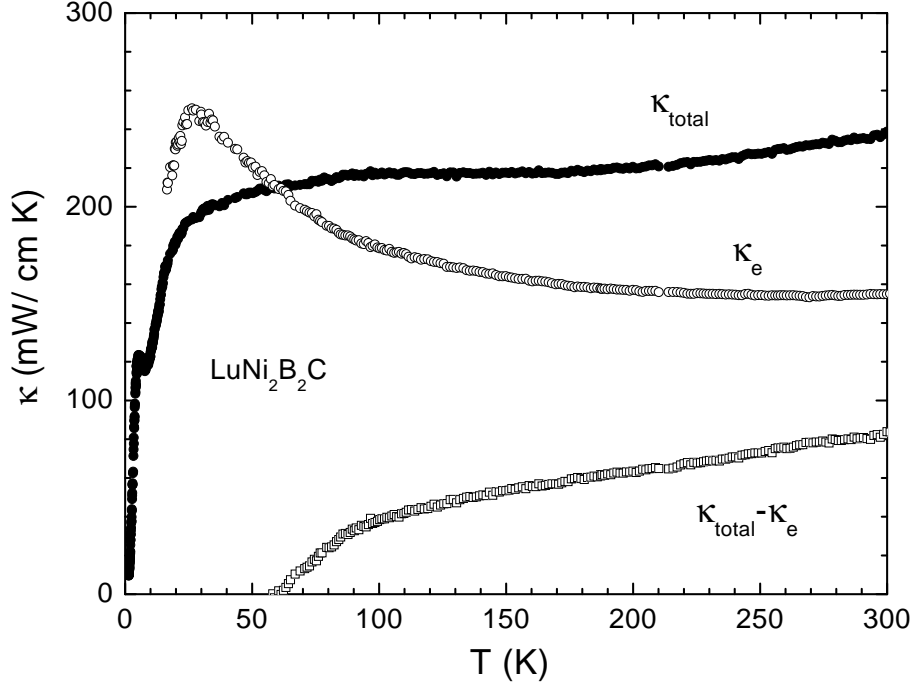,width=12cm}}
\vspace{-5pt}
\caption{{\small {\sl Temperature dependence of thermal conductivity
($\kappa_{\mathrm{total}}$) for {\rm LuNi$_2$B$_2$C}. The meaning of
 $\kappa_{\mathrm{e}}$ and $\kappa_{\mathrm{total}}$--$\kappa_{\mathrm{e}}$ is 
explained in the main text of the article. }}} 
\vspace{5pt}
\label{lutot}
\end{figure}

\par
\begin{figure}[tbh]
\centerline{\epsfig{file=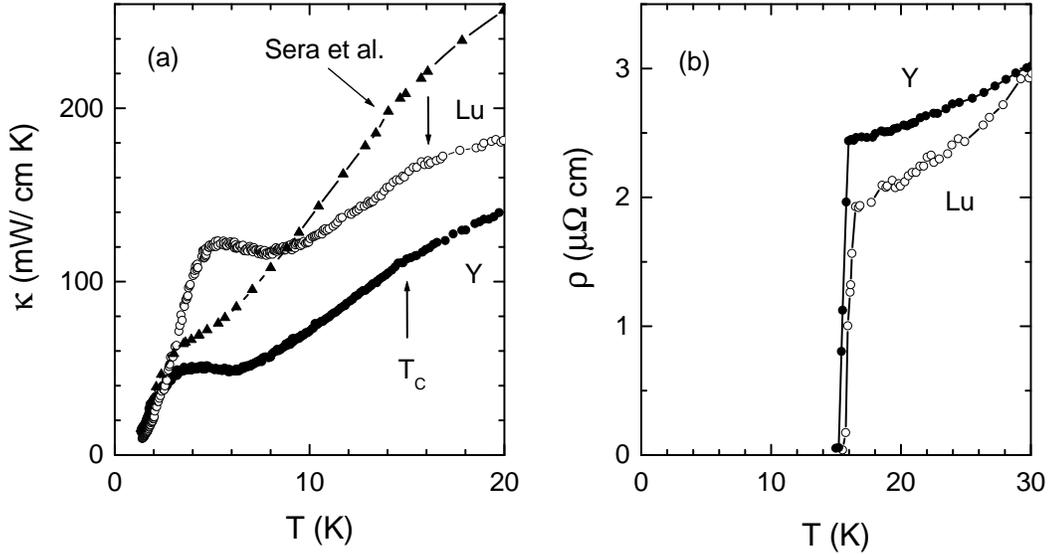,width=14cm}}
\caption{{\small {\sl $\kappa(T)$ {\rm (a)} and $\rho(T)$ {\rm (b)} 
dependences for  {\rm LuNi$_2$B$_2$C} ($\rho_{\mathrm{RT}}/\rho_{0} = 24.9$) 
and {\rm YNi$_2$B$_2$C} ($\rho_{\mathrm{RT}}/\rho_{0} = 16.5$) at 
low temperature. The left panel (a) shows also the $\kappa(T)$ data for 
{\rm YNi$_2$B$_2$C} obtained by Sera et al. \cite{sera}.}}} 
\label{yludiff}
\end{figure}

The general behavior of $\kappa(T)$ for LuNi$_2$B$_2$C is shown in 
Fig.~\ref{lutot}. Nearly the same picture is  found for YNi$_2$B$_2$C 
(and not shown in this paper for this reason). The low temperature behavior 
of $\kappa(T)$ and $\rho(T)$ is presented in Fig.~\ref{yludiff} for both 
compounds. Figure~\ref{lutot} indicates that at temperature below 
$T\simeq 60$~K the WF estimate for the electronic contribution, 
$\kappa_{\mathrm{e}}$, is greater than the total measured thermal 
conductivity.\footnote{The same behavior is found for  YNi$_2$B$_2$C for 
$T\leq 80$~K.} Obviously this is caused by the breakdown of the WF law in the 
intermediate temperature range due to inelastic electron-phonon scattering 
(see Sec.~\ref{subetc}). Really, the effective Lorenz number, $L$, in this 
temperature range could be much less than $L_{0}$, used for calculation of the 
$\kappa_{\mathrm{e}}$ (see Eq.(\ref{wf}) and Refs. \cite{parrot,ziman}). The 
study of $\kappa(T)$ of a single-crystal LuNi$_2$B$_2$C by Boaknin 
{\it et al.} \cite{boak} also implies reduced values of $L$ below 100 K. The 
deviations from the WF law are less for metals with increased lattice 
(or spin) disorder (see Sec. \ref{subetc}). Perhaps for this reason the 
breakdown of the WF law does not manifest itself so clearly in the magnetic 
borocarbides, which have higher resistivity in this temperature range than 
these of YNi$_2$B$_2$C and LuNi$_2$B$_2$C. It is interesting, however, that
the $\kappa_{\mathrm{e}}(T)$ behavior of LuNi$_2$B$_2$C (Fig. \ref{lutot}) 
corresponds to one typical of electronic thermal conductivity for fairly good 
metals, and, among other things, shows a maximum at 
$T_{\mathrm{max}}\simeq 25$~K which is expected at 
$T\leq 0.1\Theta_{\mathrm{D}}$ (see Sec. \ref{subetc}). ($\Theta_{\mathrm{D}}$ 
is about 345~K for LuNi$_2$B$_2$C \cite{carter}). This maximum (which was 
found for YNi$_2$B$_2$C as well) gives further evidence that these samples 
are well ordered. 

\par
The superconducting transitions for  LuNi$_2$B$_2$C and YNi$_2$B$_2$C are 
rather clearly indicated in $\kappa(T)$  [Fig. \ref{yludiff}(a)]. Both of the 
compounds show a strong enhancement in $\kappa$ below $T_{\mathrm{C}}$. The 
evident explanation is an increase in the phonon thermal conductivity due to 
reduced phonon-electron scattering as the normal electrons condense into 
Cooper pairs (see Sec. \ref{super}). The same behavior [that is a 
phonon-induced peak in the $\kappa(T)$ below $T_{\mathrm{C}}$] was observed by 
Boaknin {\it et al.} for LuNi$_2$B$_2$C \cite{boak}; whereas Sera {\it et al.} 
\cite {sera} have found only a ``shoulder'' in $\kappa(T)$ below 
$T_{\mathrm{C}}$ for  YNi$_2$B$_2$C which can be easily supressed by an 
external magnetic field [Sera {\it et al.} data for zero field are shown in 
Fig. \ref{yludiff}(a)]. It is obvious that the peak magnitude and its position 
depends on the relative importance of other scattering mechanisms. In fact, 
the $\kappa(T)$ curves in Fig. \ref{yludiff}(a) correspond to the case 
considered in Sec. \ref{super} when both kinds of the heat carriers (electrons 
and phonons) compete on equal terms [see Fig.~\ref{figbrt}(b)]. Taking into
account Eqs. (\ref{etc}) and (\ref{esh}),  the ratio of electronic thermal 
conductivities  in superconducting and normal states, 
$\kappa_{\mathrm{e}}^{\mathrm{s}}/\kappa_{\mathrm{e}}^{\mathrm{n}}$, below 
$T_{\mathrm{C}}$ can be presented as 
$$\kappa_{\mathrm{e}}^{\mathrm{s}}/\kappa_{\mathrm{e}}^{\mathrm{n}} = 
n_{\mathrm{n}}(T)/n_{\mathrm{e}},$$ where $n_{\mathrm{e}}$ is the electron
density, $n_{\mathrm{n}}(T)$ is the density of normal electrons below 
$T_{\mathrm{C}}$ which decreases as temperature goes down away from
$T_{\mathrm{C}}$. In the same way [assuming domination of the phonon-electron 
scattering and using Eq. (\ref{pei})] the following relation can be written 
for the ratio of phonon thermal conductivities in the superconducting and 
normal states below $T_{\mathrm{C}}$: 
$$\kappa_{\mathrm{p}}^{\mathrm{s}}/\kappa_{\mathrm{p}}^{\mathrm{n}} = 
n_{\mathrm{e}}/n_{\mathrm{n}}(T).$$ 
It is seen that a decrease in $n_{\mathrm{n}}$ below $T_{\mathrm{C}}$ leads to 
continuous reduction in the ratio 
$\kappa_{\mathrm{e}}^{\mathrm{s}}/\kappa_{\mathrm{e}}^{\mathrm{n}}$ 
for the electronic contribution; whereas, the ratio 
$\kappa_{\mathrm{p}}^{\mathrm{s}}/\kappa_{\mathrm{p}}^{\mathrm{n}}$ for the
phonon contribution rises, causing an increase in the phonon thermal 
conductivity  (see Sec. \ref{super}). The competition of these  two 
mechanisms can result in the $\kappa(T)$ behavior shown in 
Fig.~\ref{yludiff}(a). 
\par
Measurement of the thermal conductivity at very low temperatures in an 
applied magnetic field ($H_{\mathrm{c1}} \leq H \leq H_{\mathrm{c2}}$) 
provides a very powerful tool for study of the symmetry of the superconducting 
gap function. At very low temperature it is somewhat easier to sort out the 
contributions to $\kappa(T)$ from the charge carriers and the phonons. In an 
ordinary BCS type II superconductor the normal charge carriers are localized 
in the vortex cores and can only tunnel between them in the mixed state. Thus 
the electronic thermal conductivity is expected to increase very slowly at low 
field. But as the field approaches $H_{\mathrm{c2}}$, the vortices are much 
closer. The tunneling, then, becomes more pronounced and the
electronic contribution increases rapidly with applied field (see general 
description of the magnetic-field effects in the thermal conductivity below 
$T_{\mathrm{C}}$ in Ref. \cite{uher}). If,
however, there are nodes in the gap function, $\kappa(T,H)$ increases rapidly 
for even small fields above $H_{\mathrm{c1}}$, since there will be delocalized
normal electrons to couple the vortex cores, even at very low temperature.
Recent thermal conductivity measurements at low temperature for single 
crystal LuNi$_2$B$_2$C \cite{boak2} and YNi$_2$B$_2$C \cite{izawa} samples 
exhibit dramatically enhanced thermal conductivity at rather small applied 
fields. Both experiments suggest a highly anisotropic gap function for these 
borocarbides. For LuNi$_2$B$_2$C the data appear to suggest the possibility of
a line of nodes in the gap; whereas, the experiment with YNi$_2$B$_2$C has 
been interpreted in terms of point nodes along the $a$ and $b$ axes. These 
experiments certainly challenge the view that these materials are $s$-wave 
BCS superconductors. 

\vspace{25pt}
{\bf {\footnotesize c) Purely magnetic borocarbides}}
\vspace{15pt}

\par
\begin{figure}[tbh]
\vspace{-30pt}
\centerline{\epsfig{file=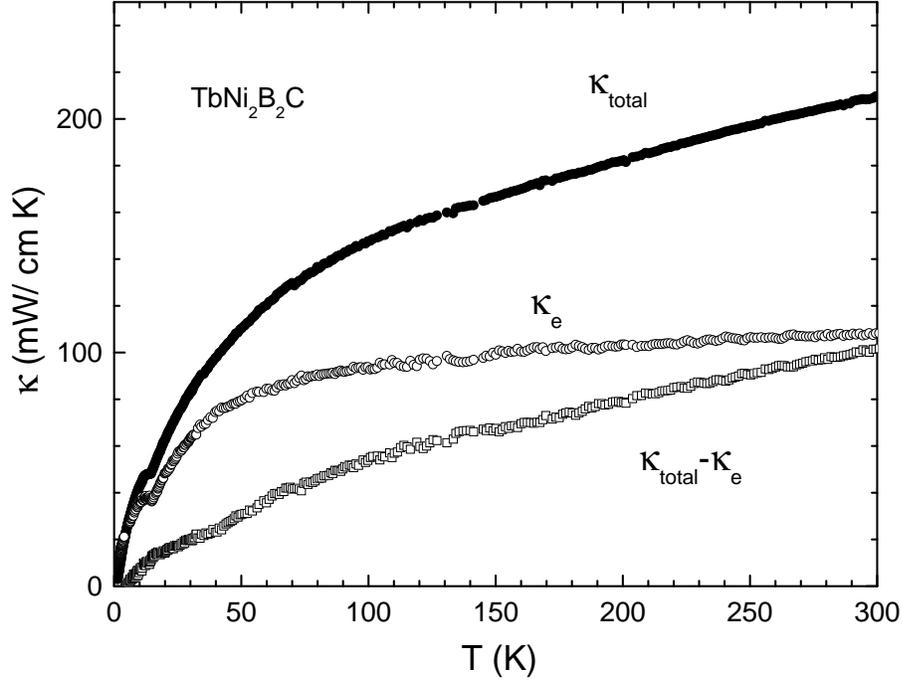,width=12cm}}
\caption{{\small {\sl Temperature dependence of thermal conductivity
($\kappa_{\mathrm{total}}$) for {\rm TbNi$_2$B$_2$C}. The meaning of
 $\kappa_{\mathrm{e}}$ and $\kappa_{\mathrm{total}}$--$\kappa_{\mathrm{e}}$ is 
explained in the main text of the article. }}} 
\label{tbtot}
\end{figure}
\begin{figure}[tbh]
\vspace{-30pt}
\centerline{\epsfig{file=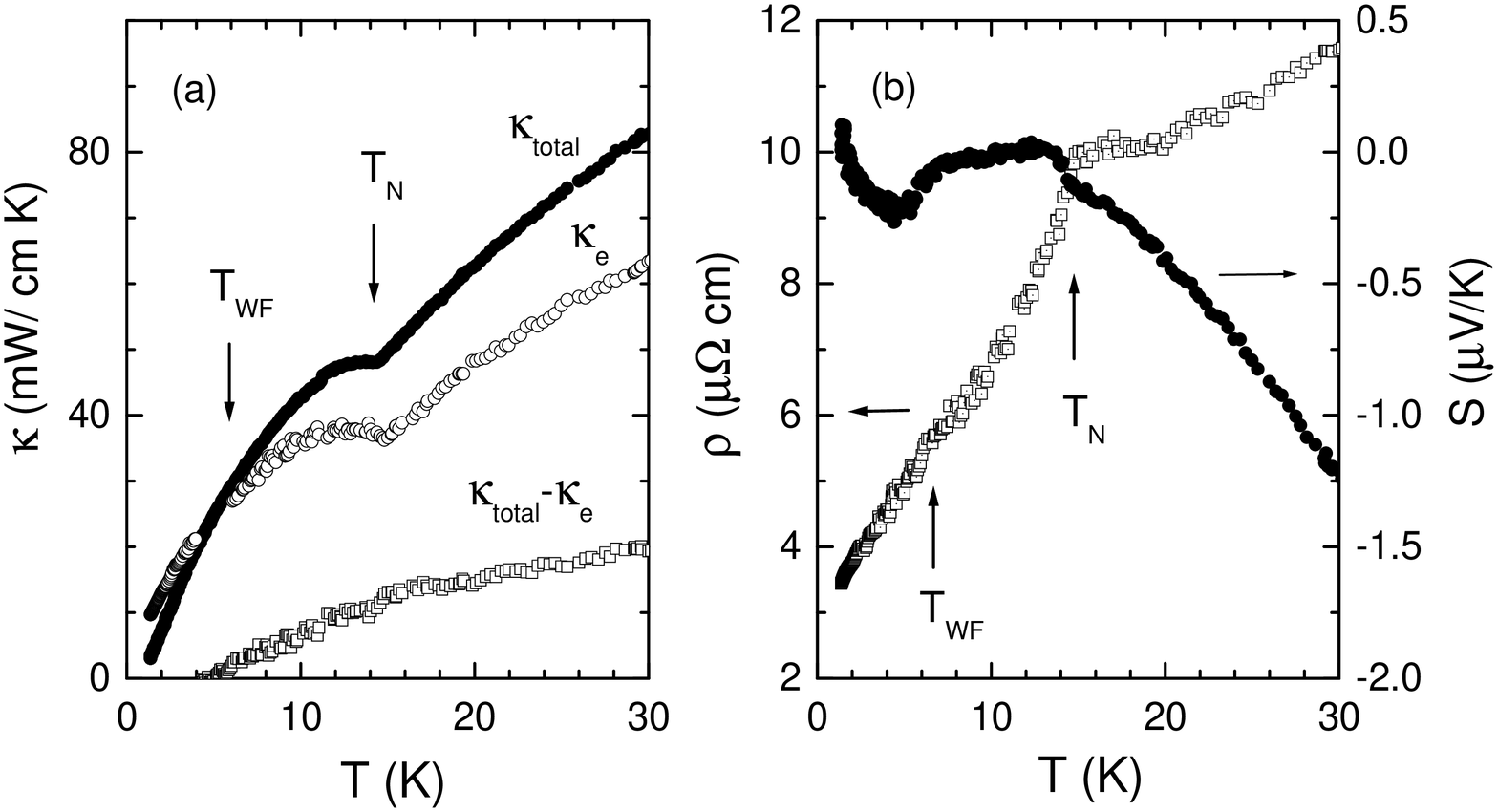,width=14cm}}
\vspace{-5pt}
\caption{{\small {\sl $\kappa(T)$ {\rm (a)}, $\rho(T)$ and $S(T)$ 
{\rm (b)} for  {\rm TbNi$_2$B$_2$C} at low temperature. The magnetic 
residual resistivity ratio, $\rho(300${\rm K})/$\rho(T_{\mathrm{N}})$ is 6.8.
Using $\rho$ at the lowest temperature measured,  
$\rho_{\mathrm{RT}}/\rho_{0} = 19.7$ can be obtained.}}} 
\label{tbdiff}
\end{figure}
\begin{figure}[tbh]
\vspace{-30pt}
\centerline{\epsfig{file=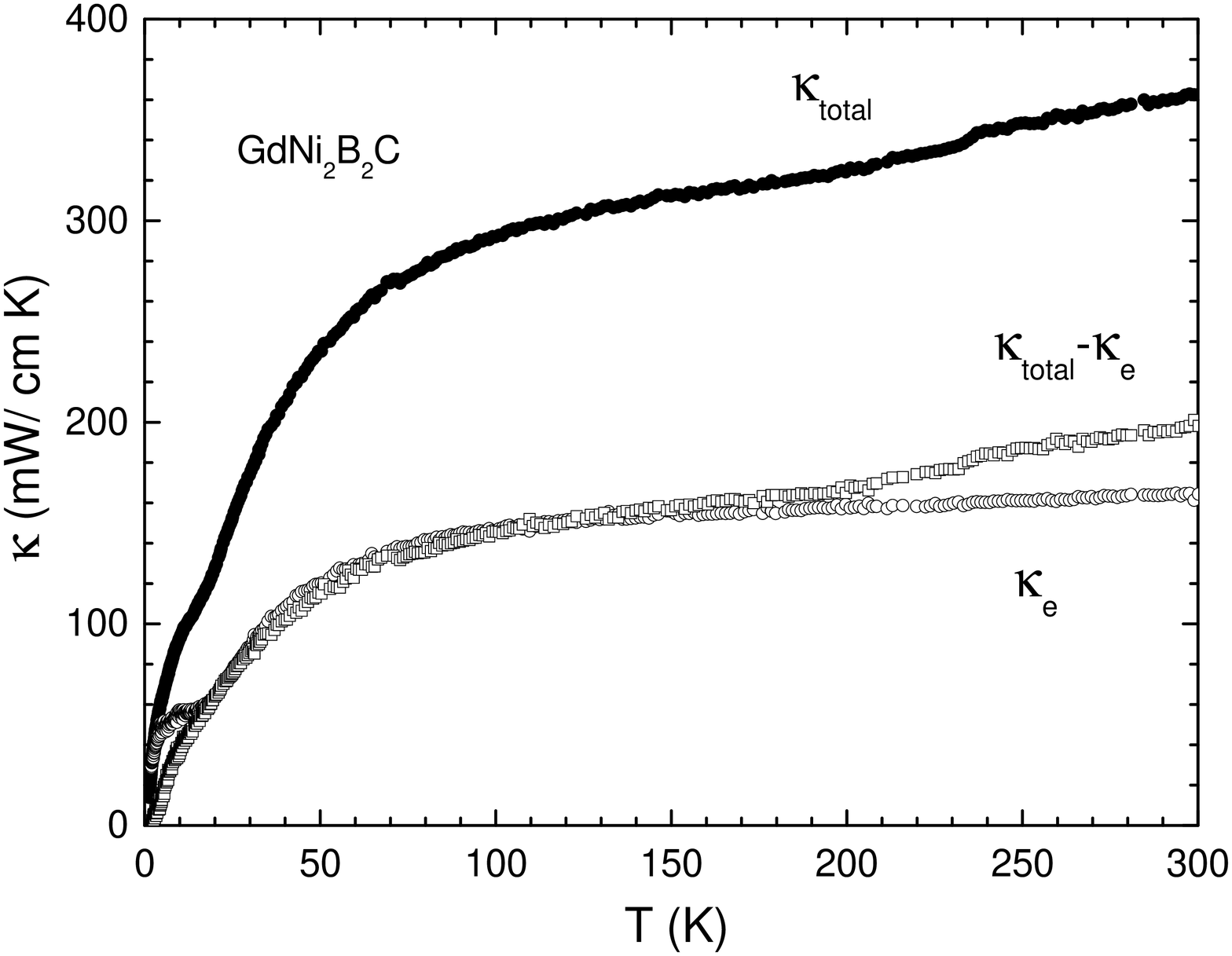,width=12cm}}
\caption{{\small {\sl Temperature dependence of thermal conductivity
($\kappa_{\mathrm{total}}$) for {\rm GdNi$_2$B$_2$C}. The meaning of
 $\kappa_{\mathrm{e}}$ and $\kappa_{\mathrm{total}}$--$\kappa_{\mathrm{e}}$ is 
explained in the main text of the article. }}} 
\label{gdtot}
\end{figure}
\begin{figure}[tbh]
\vspace{-30pt}
\centerline{\epsfig{file=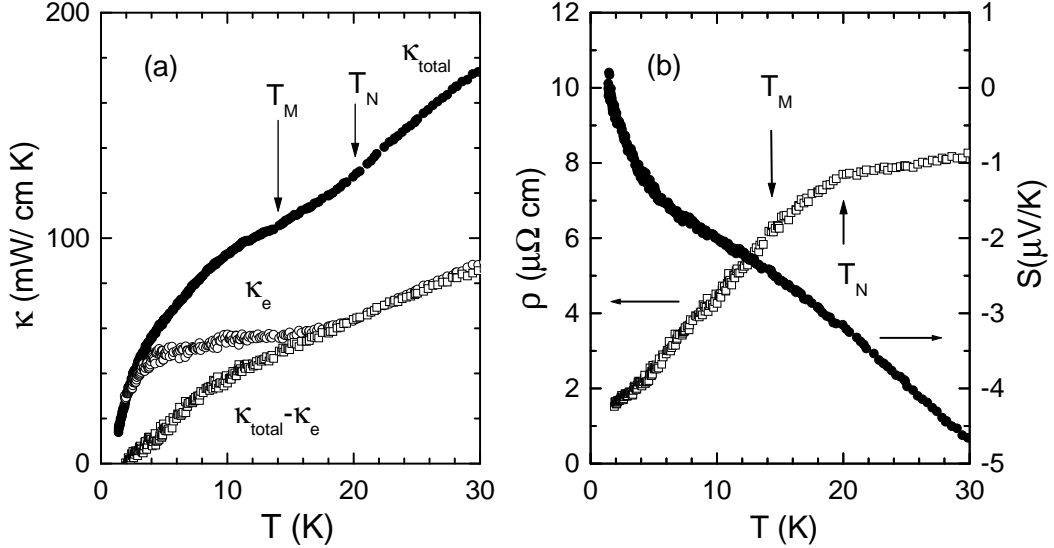,width=14cm}}
\vspace{-5pt}
\caption{{\small {\sl $\kappa(T)$ {\rm (a)}, $\rho(T)$ and $S(T)$ 
{\rm (b)} dependences for  {\rm GdNi$_2$B$_2$C} at low temperature.
The magnetic residual resistivity ratio, 
$\rho(300${\rm K})/$\rho(T_{\mathrm{N}})$ is 5.8. Using $\rho$ at the lowest 
temperature measured,  $\rho_{\mathrm{RT}}/\rho_{0} = 29.7$ can be obtained.
}}} 
\label{gddiff}
\end{figure}

Borocarbides {\bf TbNi$_2$B$_2$C} and {\bf GdNi$_2$B$_2$C} 
are magnetic at low enough temperature (see Table I), but not superconducting
(at least above 0.3 K). The absence of superconductivity is supposed to be
caused by 4$f$ magnetic moments \cite{muller}. The temperature behavior of 
thermal conductivity for these compounds is presented in Figs. \ref{tbtot} and 
\ref{gdtot}. Expanded views of $\kappa(T)$, $\rho(T)$ and $S(T)$ dependences 
at low temperature are shown in Figs. \ref{tbdiff} and \ref{gddiff}. 
\par
Two magnetic transitions have been found for TbNi$_2$B$_2$C: (1) transition 
(of first order) 
\newpage

\newpage 

\newpage
\noindent 
from paramagnetic to modulated AFM state below $T_{\mathrm{N}}\approx 15$~K 
with the magnetic moments lying along the $a$ 
direction, and (2) transition (of second order) to WFM state below 
$T_{\mathrm{WF}}=$6--8 K with the magnetic moments lying along the $<\!100\!>$ 
or $<\!110\!>$ directions in the $ab$ plane under applied  magneic fields 
\cite{lynn,cho,tb}. The transition to the modulated AFM state is
clearly indicated in the $\kappa(T)$, and causes strong changes in $\rho(T)$ 
and $S(T)$ curves (Fig. \ref{tbdiff}). This can be explained in the same way 
as for DyNi$_2$B$_2$C and HoNi$_2$B$_2$C by an appreciable decrease  in the
rate of electron scattering by the spin disorder in response to PM-AFM
transition (see discussion above). Since in this compound the electronic 
contribution to the thermal conductivity, $\kappa_{\mathrm{e}}$, clearly 
dominates over the phonon one, 
$\kappa_{\mathrm{total}}$--$\kappa_{\mathrm{e}}$, for the low temperature
range (see Fig. \ref{tbtot}), the measured thermal conductivity, 
$\kappa_{\mathrm{total}}$, can exhibit a clear respons to this change in the 
electron scattering rate. The second-order AFM-WFM transition at
$T_{\mathrm{WF}}$ is not indicated in $\kappa(T)$ dependence, although weak
change in $\rho(T)$  and rather appreciable change in $S(T)$ (both around 
$T\approx 7$~K) are clearly seen (Fig. \ref{tbdiff}). The absence of some 
feature in $\kappa(T)$ at $T_{\mathrm{WF}}$ can be explained by the fact that 
this transition does not strongly effect the scattering of charge carriers 
and has no influence on the phonon scattering. 
\par
The  borocarbide GdNi$_2$B$_2$C transforms into modulated AFM state below 
$T_{\mathrm{N}}\approx 20$~K (first-order transition), with the magnetic 
moments lying in the $b$ direction \cite{gd1,gd2}. With further decreasing 
temperature it goes into another modulated AFM state below 
$T_{\mathrm{M}}\approx 13.6$~K (second-order transition), with the magnetic 
moments tilting somewhat into the $c$ 
direction \cite{gd1,gd2}. The measured $\kappa_{\mathrm{total}}(T)$ dependence 
shows a subtle change in its slope at $T_{\mathrm{N}}$ (Fig. \ref{gddiff}), 
which is less clear than that for TbNi$_2$B$_2$C (Fig. \ref{tbdiff}). The weak 
feature in $\kappa(T)$ at $T_{\mathrm{N}}$ for GdNi$_2$B$_2$C was found also 
for polycrystalline samples in Ref. \cite{cao}. 
This weak sensitivity of $\kappa(T)$ to first-order AFM transition
can be partly explained by the fact that the electronic contribution to 
total $\kappa$ for this compound is much less than that of in TbNi$_2$B$_2$C 
(compare Figs. \ref{tbtot} and \ref{gdtot}) and (especially) in DyNi$_2$B$_2$C 
and HoNi$_2$B$_2$C (see Figs. \ref{dytot} and \ref{hotot}), where the magnetic 
transitions of this type produce quite appreciable changes in $\kappa(T)$
at $T_{\mathrm{N}}$ (Figs. \ref{dydiff} and \ref{hodiff}). The 
transition to another MAFM state below $T_{\mathrm{M}}\approx 14$~K produces
only subtle changes in $\kappa(T)$, $\rho(T)$ and $S(T)$ of GdNi$_2$B$_2$C
(Fig.~\ref{gddiff}). The explanation of this fact is the same as indicated 
above for TbNi$_2$B$_2$C, i.e., the transition appears to only slightly 
affect the scattering of charge carriers and has no influence on the phonon 
scattering.  It is interesting to note that here is no indication of 
$T_{\mathrm{N}}$ in the phonon contribution for R = Tb or Gd, unlike the small 
dip at $T_{\mathrm{N}}$ in the phonon contribution for R = Dy shown in in 
Fig. \ref{dydiff}. This suggests that the nature of the transitions for the 
former two materials is different in regards to its influence on the phonon 
contributions. An alternate explanation is that the small dip in Fig. 
\ref{dydiff} for DyNi$_2$B$_2$C is an artifact of using the WF law in the 
region where $\rho(T)$ is varying so rapidly with temperature. Any small 
variations in temperature between the $\rho(T)$ and $\kappa(T)$ measurements 
would create this dip.

\vspace{25pt}
{\bf {\footnotesize d) Heavy-fermion borocarbide 
YbNi$_2$B$_2$C}}
\vspace{15pt}
\par
{\bf YbNi$_2$B$_2$C} is neither superconducting (above 0.05 K \cite{cyb}) nor 
orders magnetically (above 0.023 K \cite{bon}). The compound is a 
heavy-fermion system \cite{ayb,bon,byb} with a Kondo temperature of 
$T_{\mathrm{K}}\approx 10$~K and with a Sommerfeld coefficicent of 
$\gamma\simeq 530$~mJ/(mole~K$^{2}$) \cite{ayb}. When compared to that 
of LuNi$_2$B$_2$C ($\gamma \approx 19$~mJ/(mole K$^{2})$ \cite{carter,michor})
this corresponds to an effective mass almost 30 times larger than that for
the Lu-based compound provided that $k_{\mathrm{F}}$ is approximately the same 
for all members of the series. The magnetic susceptibility is anisotropic and 
exhibits Curie-Weiss [$1/(T-\theta_{\mathrm{CW}})$] behavior at high 
temperature with a negative value of $\theta_{\mathrm{CW}}$, indicative  
that AFM correlations play a significant role. This behavior is consistent 
with the interpretation of a highly correlated ground state at low 
temperatures and crystal electric field effect at higher temperatures 
\cite{cyb}. Recent neutron scattering experiments of Boothroyd {\it et al.} 
\cite{booth} indicate that the crystal electric field is significantly 
enhanced over that of the other family members and produces a temperature 
dependent effective Kondo interaction. These authors found a Kondo temperature 
of 25 K compared to the $T_{\mathrm{K}} = 10$~K, obtained from specific heat
and susceptibility measurements \cite{ayb}, and they also suggest that at 
$T=0$ the compound might be close to a quantum critical point on the 
non-magnetic side. The electrical 
resistivity exhibits a quadratic temperature dependence below 1.5 K 
\cite{ayb}, and the ratio of the coefficient of the quadratic term to the 
$\gamma^{2}$ is approximately that  found for UPt$_{3}$ along its hexagonal 
axis \cite{ayb}. These results all suggest that hybridization between the 
Yb $4f$ and conduction electron states is responsible for the suppression of 
superconductivity in this borocarbide \cite{ayb} even though a simple de 
Gennes scaling argument \cite{ayb,byb} would suggest that it should be 
superconducting at about 12 K and magnetically order at about 0.4 K. 

\begin{figure}[tbh]
\vspace{8pt}
\centerline{\epsfig{file=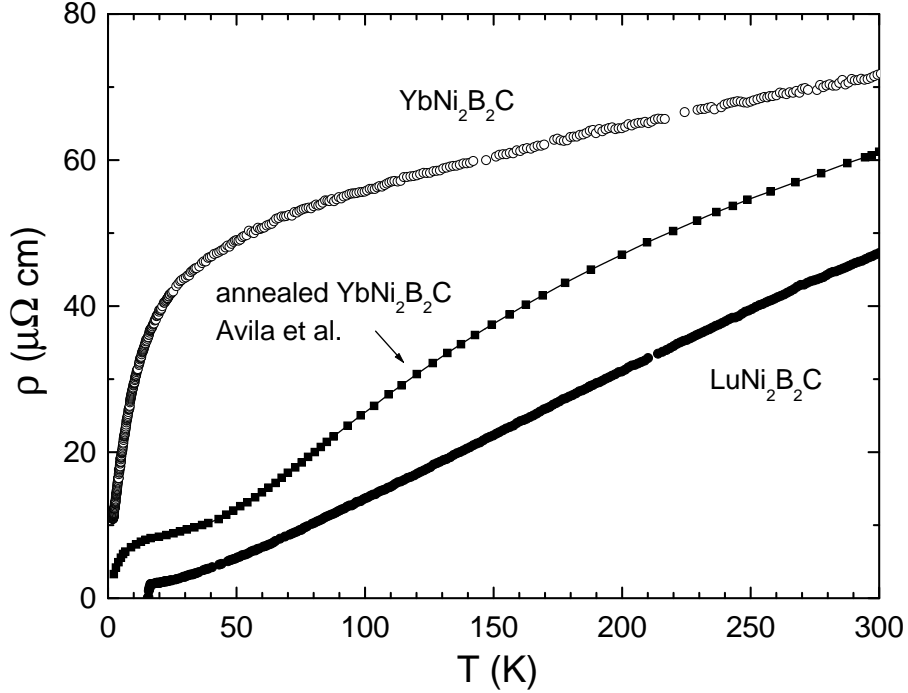,width=12cm}}
\caption{{\small {\sl Temperature dependences of resistivity for 
{\rm YbNi$_2$B$_2$C} and {\rm LuNi$_2$B$_2$C}. The data of Avila et al.
\cite{avila} show $\rho(T)$ for a single crystal {\rm YbNi$_2$B$_2$C} annealed 
for 150 hours at 950$^{\circ}${\rm{C}}.}}} 
\label{ybrho}
\end{figure}

\par
Single crystal alloy samples Lu$_{1-x}$Yb$_{x}$Ni$_2$B$_2$C have been used to
explore the transition from superconductivity through single-impurity Kondo 
behavior to Kondo lattice heavy-fermion behavior from $x=0$ to $x=1$ 
through transport (resistivity and thermopower) \cite{fyb,gyb} and 
thermodynamic (susceptibility and specific heat) measurements \cite{gyb,hyb}.
The suppression of $T_{\mathrm{C}}$ with Yb substitution does not scale with
the de Gennes factor as would be expected from the Abrikosov-Gor'kov theory
of magnetic pair breaking \cite{eyb}. For example, the suppression rate of 
$T_{\mathrm{C}}$ at small concentrations $x$ is 75 times higher for Yb than 
for Gd \cite{iyb}. Recent studies \cite{gyb,hyb} indicate that the Kondo 
temperature varies strongly with Yb concentration $x$ and that the rapid 
suppression of $T_{\mathrm{C}}$ with Yb concentration is consistent with 
M\"{u}ller-Hartmann and Zittartz theory \cite{jyb} for superconducting Kondo 
systems with spin 1/2 and $T_{\mathrm{K}}/T_{\mathrm{C}} \approx 10^3$.

\begin{figure}[tbh]
\vspace{8pt}
\centerline{\epsfig{file=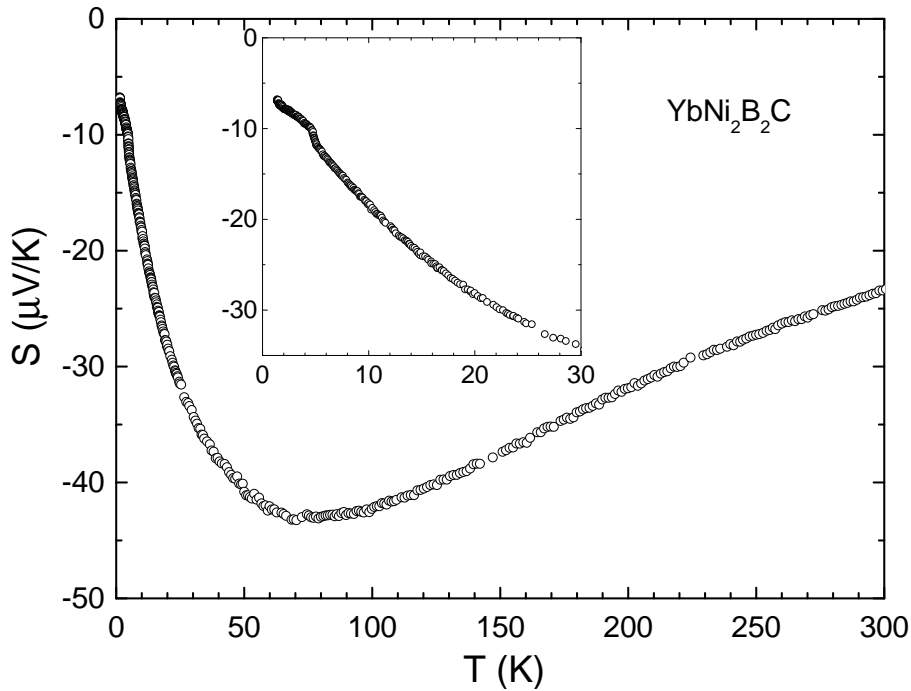,width=12cm}}
\caption{{\small {\sl Temperature dependence of thermo-electric power for 
{\rm YbNi$_2$B$_2$C}.}}} 
\label{ybs}
\end{figure}

\par
\begin{figure}[tbh]
\vspace{8pt}
\centerline{\epsfig{file=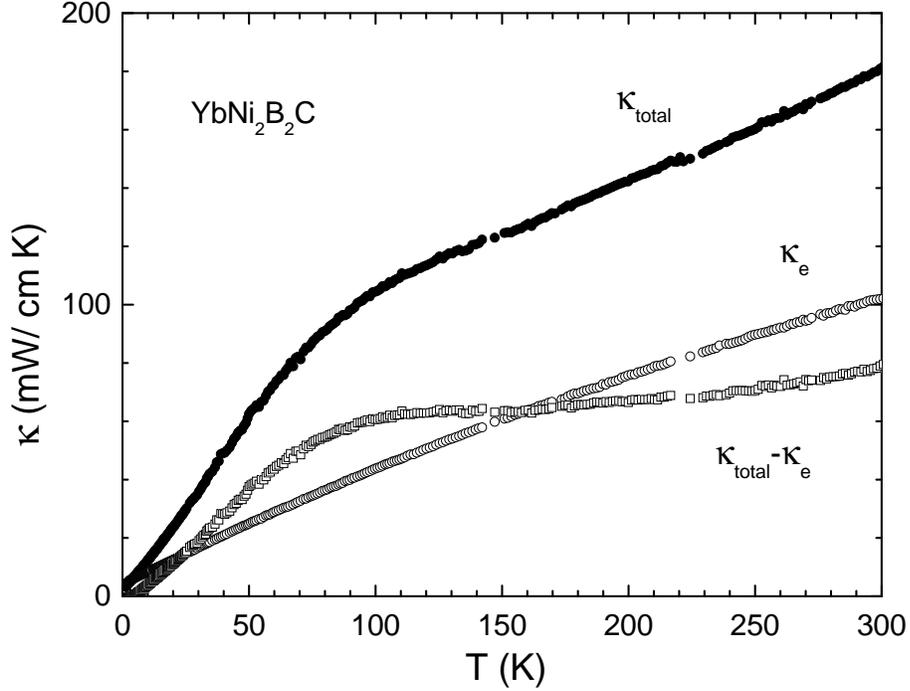,width=12cm}}
\caption{{\small {\sl Temperature dependence of thermal conductivity
($\kappa_{\mathrm{total}}$) for {\rm YbNi$_2$B$_2$C}. The meaning of
 $\kappa_{\mathrm{e}}$ and $\kappa_{\mathrm{total}}$--$\kappa_{\mathrm{e}}$ is 
explained in the main text of the article. }}} 
\vspace{10pt}
\label{ybtot}
\end{figure}

Although thermopower and resistivity measurements have been reported for the
Lu--Yb borocarbide alloys, the thermal conductivity has only been reported
\cite{brian2} for YbNi$_2$B$_2$C. To illustrate the behavior of this system, 
the $ab$-plane resistivity, thermopower and thermal conductivity taken from 
Ref. \cite{brian2} for an unannealed  single crystal sample are shown in 
Figs. \ref{ybrho}, \ref{ybs}, and \ref{ybtot}, respectively. The resistivity
of a single crystal sample of LuNi$_2$B$_2$C and that of the annealed 
YbNi$_2$B$_2$C sample \cite{avila} are also shown in Fig. \ref{ybrho} 
for comparison. At high temperature the resistivity varies approximately 
linearly with temperature and is large in magnitude compared to most metals. 
The magnitude of the room-temperature resistivity of this sample is about 
one-fourth of that for the polycrystalline sample in Ref. \cite{byb} and 
one-half of that for the single crystal sample in Ref. \cite{ayb}. The large 
variations in the room temperature resistivity suggest unusual disorder 
effects. The room temperature resistivity for these unannealed Yb-based 
samples are all much larger than that for R = Lu, or even for the magnetic 
borocarbides of this family. Recent measurements \cite{avila} show that, 
although the thermodynamic properties are independent of the degree of 
disorder for this heavy fermion compound, the transport properties are very 
sensitive. Annealing the single crystals for 150 hours at 950$^{\circ}$C 
reduced the room temperature resistivity for one sample from almost 
100 $\mu\Omega$~cm to 60 $\mu\Omega$~cm and also altered the temperature 
dependence $\rho(T)$ appreciably (Fig. \ref{ybrho}). The beginning of a Kondo 
minimum was observed at about 30 K followed by the very sharp decrease in 
resistance due to coherence around $T_{\mathrm{K}}$. Since disorder around the 
hybridizing Yb site can greatly change the ``local'' Kondo temperature, these 
authors interpret the large resistivity and different temperature dependence 
of $\rho(T)$ as due to a distribution of local Kondo temperatures 
\cite{avila}. They estimate that only a few percent of sites need to have  
$T_{\mathrm{K}}$ much greater than 10 K to explain these differences.

\par
The in-plane thermopower $S(T)$ in this unannealed single crystal is negative 
at all temperatures and very large in magnitude compared to the other 
compounds (see Fig. \ref{ybs}). It decreases linearly with decreasing 
temperature near room temperature and shows a minimum of about -44 $\mu$V/K 
near 70 K. At low temperature a distinct shoulder can be seen near 5 K in the 
insert. This behavior is quite different from that found in the other 
rare-earth nickel borocarbide family members \cite{don2}. These data are in 
general agreement with other measurements of $S(T)$ for unannealed single 
crystals of YbNi$_2$B$_2$C \cite{don,fyb,gyb}. The magnitude of thermopower, 
however, is also 
affected rather strongly by disorder, but the temperature dependence is not 
much affected. 
Avila {\it et al.} \cite{avila} report that annealing a single crystal for 150 
hours at 950$^{\circ}$C reduces the magnitude of $S(T)$ at the minimum from 
about 40$\mu$V/K to about 20 $\mu$V/K and moves the minimum to somewhat higher 
temperature. The influence of disorder at lower temperatures has not been 
determined since they have measured $S(T)$ only for $T\geq 100$~K. 
\par
The thermal conductivity of this unannealed heavy-fermion sample is shown in 
Fig.~\ref{ybtot}. It is rather featureless except for the rapid decrease with 
decreasing temperature that starts around 100 K. Because of the large 
resistivity at high temperature the WF law is expected to be valid over much 
of this temperature range. The phonon contribution inferred from the WF law
($\kappa_{\mathrm{total}}$--$\kappa_{\mathrm{e}}$) is appreciably smaller at
high temperature than that indicated for the other borocarbide samples 
discussed above, but a similar increase with temperature at high $T$ is 
observed for nearly all of the samples. As discussed earlier, this is most 
likely due to the dominance of PI and/or PE scattering. Although there is no 
$\kappa(T)$ data available for well annealed single crystal, one should expect 
singificantly different behavior for the electronic contribution due to the 
different $\rho(T)$ dependence reported \cite{avila}.

\subsection{Rutheno-Cuprates}
Coexistence of superconductivity with ferromagnetism (FM) is one of the 
fundamental problems in condensed matter physics that has been studied over 
several decades.  Recently, however, coexistence of these two mutually 
exclusive phenomena have been observed in layered cuprate systems where 
magnetic order first sets in at temperatures $T_{\mathrm{M}}$ as high as about 
180 K and superconductivity as high as about 50 K.  The ratio 
$T_{\mathrm{M}}/T_{\mathrm{C}}\approx 4$ is much greater than that described 
in the intermetallic compounds discussed in sections \ref{int} or \ref{boro}. 
(See Felner \cite{a} and Lorenz, Xue and Chu \cite{lorenz} for a recent
 review.)  A particularly interesting system where coexistence of weak 
FM with superconductivity has been reported is 
(R$_{1+x}$Ce$_{1-x})$RuSr$_2$Cu$_2$O$_{10-\delta}$ (Ru-2122 phase), where 
R = Eu or Gd.  The compounds, analogous to the Nb-2122 system \cite{b}, are 
related to the YBa$_2$Cu$_3$O$_7$ (123) structure with a fluorite type 
(R$_{1+x}$Ce$_{1-x}$)O$_2$ layer separating the CuO$_2$ double layers instead 
of the rare earth Y-layer and with RuO$_6$ octahedra replacing the CuO$_4$ 
squares in the CuO plane for the 123 high-$T_{\mathrm{C}}$ compound \cite{c}.  
Because of this fluorite type layer, the successive perovskite blocks are 
shifted by half a diagonal of the $ab$-plane unit cell and the cell is 
sufficiently long to encompass two formula 
units per unit cell. The crystal structure is shown in Fig. 1 of the 
accompanying article by Felner \cite{a} and Fig. 3 in the accompanying 
article by Lorenz {\it et al.} \cite{lorenz}.  The Ru ion is pentavalent in 
this compound.  Superconductivity is thought to reside primarily in the 
CuO$_2$ planes, and the magnetic behavior appears to be associated with the Ru 
sites.  Coexistence of superconductivity and weak FM was first 
discovered in the Ru-2122 phase for R = Eu and Gd \cite{d,e}.  It was also 
reported for the Ru-1212 phase compound GdSr$_2$RuCu$_2$O$_8$ \cite{f}.
\par
The only thermal conductivity measurements reported for this system are for a 
Ru-2122 sample with $x=0.5$ and R = Eu which has been annealed in pure oxygen 
at 54 atm. to provide the hole doping in the CuO$_2$ planes \cite{brian1}.  
Consequently, we will focus on coexistence of weak FM and superconductivity in 
the compound 
(Eu$_{1+x}$Ce$_{1-x}$)RuSr$_2$Cu$_2$O$_{10-\delta}$ with $x=0.5$ as evidenced 
by thermopower and thermal conductivity.  A detailed review of the other 
properties of the Ru-2122 phase compounds is given by Felner \cite{a} and 
by Lorenz {\it et al.} \cite{lorenz}.  They will be only briefly summarized 
here before discussion of the thermopower and thermal conductivity.
\par
Hole-doping in this system of compounds can be accomplished by either 
adjusting the R$^{3+}$/Ce$^{4+}$ ratio or the oxygen content.  The parent 
insulator compound is assumed to be that for $\delta=0$, $x=0$ \cite{a}.  
The formal valences then associated with this compound would be 
Eu$^{3+}$Ce$^{4+}$Ru$^{5+}$[Sr$^{2+}$]$_{2}$[Cu$^{2+}$]$_2$[O$^{2-}$]$_{10}$.  
Superconductivity is observed for $0.2\leq x\leq 0.6$ with the optimally doped 
sample corresponding to $x=0.4$.  The variation of $T_{\mathrm{C}}$ over this 
range of $x$ is only a few Kelvin below that of the optimal value 
$T_{\mathrm{C}} = 35$~K (for $x = 0.4$) in the as prepared samples, much less 
than that expected if all of the carriers were introduced into the CuO$_2$ 
planes.  One explanation that has been suggested for this is, that the holes 
introduced in the CuO$_2$ planes by replacing Ce$^{4+}$ with R$^{3+}$ are 
partially compensated by a deficiency $\delta$ in the oxygen content 
\cite{a}.  Neutron diffraction measurements for 
Gd$_{1.3}$Ce$_{0.7}$RuSr$_2$Cu$_2$O$_{10-\delta}$ indicate $\delta=0.22$ with 
the deficiency primarily in the fluorite layer and RuO$_2$ planes \cite{g}.  
Recent L$_{\mathrm{III}}$-edge XANES spectra for 
Gd$_{1+x}$Ce$_{1-x}$RuSr$_2$Cu$_2$O$_{10}$ with $x=0.9$ and 0.5, establish 
that Ru is pentavalent independent of the Ce concentration \cite{a}, 
i.e. there is no charge transfer to the RuO$_2$ layer. In this case the formal 
valence count would be 
[Gd$_{1+x}$Ce$_{1-x}$]$^{+7-x}$Ru$^{+5}$[Sr$_{2}$]$^{+4}$[Cu$_{2}$]$^{+4-x+2\delta}$[O$_{10-\delta}$]$^{-20+2\delta}$. Thus the valence for Cu in this case of 
$x=0.3$ and $\delta=0.22$ would be approximately +2 as in the case of 
GdRuSr$_2$Cu$_2$O$_8$. Knee {\it et al.} \cite{g} have suggested that a hole
doping mechanism arising from overlap of the t$_{\mathrm{2g}}$ band of Ru and
the $\mathrm{d}_{x^{2}-y^{2}}$ one of Cu, that is thought to be responsible 
for superconductivity in  GdRuSr$_2$Cu$_2$O$_8$, may also be responsible for 
superconductivity in Gd(Eu)$_{1+x}$Ce$_{1-x}$RuSr$_2$Cu$_2$O$_{10-\delta}$ 
compounds. Hole 
doping can also be accomplished by annealing the samples at high temperature 
and pressure in pure oxygen.  $T_{\mathrm{C}}$ has been raised monotonically 
from 34~K to as high as 49~K by annealing 
Eu$_{1.5}$Ce$_{0.5}$RuSr$_2$Cu$_2$O$_{10-\delta}$ at 800$^{\circ}$C in pure 
oxygen for 24 hours at pressures up to 150 atm. \cite{h}.  Unfortunately, the 
absolute oxygen content is difficult to determine in this compound.  
Consequently, neither $T_{\mathrm{C}}$ as a function of oxygen content nor the 
location of the excess oxygen in this compound has been systematically 
determined.  Also, the oxygen appears to diffuse back out of these oxygenated 
samples if held in a vacuum at room temperature.
\par
\begin{figure}[tbh]
\vspace{-20pt}
\centerline{\epsfig{file=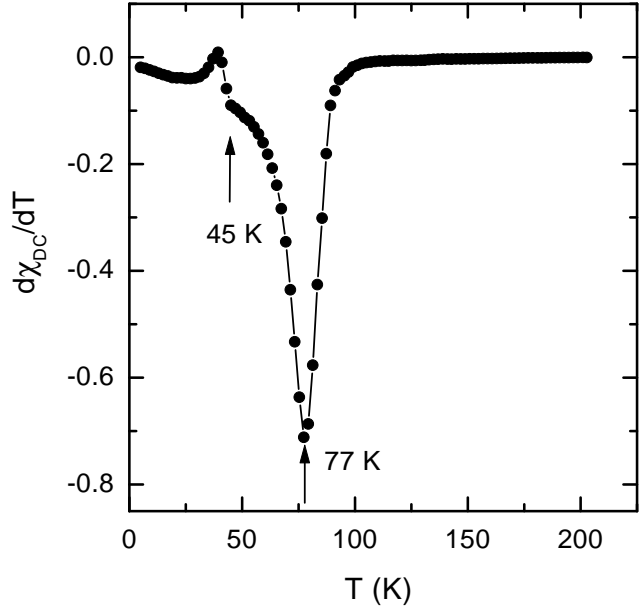,width=9cm}}
\caption{{\small {\sl 	Derivative of the field-cooled (FC) DC-susceptibility
$d\chi_{\mathrm{DC}}/dT$ [$\chi_{\mathrm{DC}}=M/H$ was recorded at
$H=50$~Oe] as a function of temperature $T$ for an oxygenated
Eu$_{1.5}$Ce$_{0.5}$RuSr$_2$Cu$_2$O$_{10-\delta}$ sample (taken from
Ref. \cite{brian1}).  The maximum value in the magnitude at about 77K, very
near the point where the zero-field-cooled (ZFC) and FC curves for
$\chi_{\mathrm{DC}}$ approximately appear to merge, is identified as the
irreversibility temperature $T_{\mathrm{irr}}$.  The onset of
superconductivity is identified as the sharp break in $d\chi_{\mathrm{DC}}/dT$
at about 45K.  The onset of antiferromagnetic order $T_{\mathrm{M}}$
determined by the appearance of non-linearity in the $M(H)$ curves at
$T_{\mathrm{M}}=180$~K does not produce an observable feature in
$d\chi_{\mathrm{DC}}/dT$.
}}}
\label{felner1}
\end{figure}

The magnetic behavior of the Eu$_{1+x}$Ce$_{1-x}$RuSr$_2$Cu$_2$O$_{10-\delta}$ 
is also quite unusual \cite{a}.  In the high temperature region, Curie-Weiss 
paramagnetism [$1/(T-\theta_{\mathrm{CW}})$ law] is observed with an 
effective moment, $g\sqrt{S(S+1)}$, of about 2.15~$\mu_{\mathrm{B}}$ for Ru 
and a value of $\theta_{\mathrm{CW}}$ in the interval between 134~K and 146~K, 
relatively independent of the Ce concentration.  This effective moment lies 
between that of the high spin state for Ru$^{5+}$ ($S=3/2$,  
3.9~$\mu_{\mathrm{B}}$) and the low-spin state ($S=1/2$,  
1.7~$\mu_{\mathrm{B}}$), and the Curie-Weiss temperature 
$\theta_{\mathrm{CW}}$ is positive, indicative of a ferromagnetic interaction. 
The low temperature saturation magnetization ($gS\mu_{\mathrm{B}}$) is only 
0.89 $\mu_{\mathrm{B}}$/Ru ion compared to the value expected for the 
high-spin (3.0 $\mu_{\mathrm{B}}$) and low-spin (1.0 $\mu_{\mathrm{B}}$) 
states for pentavalent Ru. At the temperature $T_{\mathrm{M}}$ the Ru orders 
antiferromagnetically as determined by the appearance of non-linearity of in 
$M(H)$.  At a lower temperature $T_{\mathrm{irr}}$, indicated by the 
appearance of a remnant magnetization and approximate merging of the 
field-cooled and zero-field-cooled $M(T)$ curves, weak FM and 
irreversibility is produced by canting of the Ru moments as a result of a 
Dzyaloshinsky-Moriya type antisymmetric exchange coupling \cite{i} between 
neighboring Ru moments that is induced by distortion of the RuO$_6$ 
octahedrons.  The sharp minimum at 77 K in the temperature derivative of the 
field-cooled DC susceptibility ($\chi_{\mathrm{DC}}=M/H$) shown in 
Fig.~\ref{felner1} coincides with this transition for a 
Eu$_{1.5}$Ce$_{0.5}$RuSr$_2$Cu$_2$O$_{10-\delta}$ sample that has been 
annealed at 54 atm. in pure oxygen at $800^{\circ}$C for 12 hours 
\cite{brian1}.  The sharp rise in the derivative at 45 K coincides with the 
onset of superconductivity at $T_{\mathrm{C}}$.  Below this temperature weak 
FM and superconductivity coexist, presumably on a microscopic scale.  No 
feature indicative of $T_{\mathrm{M}}$ is observed in 
$d\chi_{\mathrm{DC}}/dT$.\footnote{Note that a totally different explanation 
of the magnetic behavior for this type of compound has been suggested by 
Lorenz {\it et al.} \cite{lorenz}. They suggest that there is a mesoscopic
phase separation between the FM and the AFM species with superconductivity
existing only in the AFM grains separated by nanoscale FM domains.
Superconductivity would develop in the AFM domains which are Josephson coupled
across the FM regions. The onset of the intergrain superconductivity is then
expected to be the result of a phase-lock transition of an array of Josephson
junctions rather than coexistence of superconductivity with ferromagnetism.}
\par
Different possibilities for the structure of coupled inhomogeneous
superconducting and magnetic order parameters for weak FM  have been
previously proposed in the literature: the Fulde-Ferrell-Larkin-Ovchinnikov
state \cite{fulde} where the superconducting order parameter develops a
spatial variation with non-zero total momentum, the spiral magnetic (SM) state
\cite{j}, the spontaneous vortex (SV) state \cite{k}, the linearly polarized
(LP) state \cite{l}, the linearly oscillating vortex (LOV) state \cite{m}, and
the spiral magnetic vortex (SMV) state \cite{m}.  Sonin and Felner \cite{n}
have argued that the predicted SV phase \cite{k} is the most likely candidate
to describe this state.

\par
The temperatures $T_{\mathrm{M}}$ and $T_{\mathrm{irr}}$ depend rather
strongly on the doping, and even on the method of doping.   $T_{\mathrm{M}}$
and $T_{\mathrm{irr}}$ decrease approximately linearly with the reduction $x$
in Ce content for the as prepared samples \cite{a}.  This corresponds to
adding holes to the CuO$_2$ layer, even though they may be partially
compensated by an induced oxygen deficiency.  As previously noted, there is an
optimal doping for $x=0.4$ which gives the maximum in  $T_{\mathrm{C}}$.
If the hole doping via Ce arises from overlap of the Ru and Cu $d$-bands, the
magnetic behavior of Ru may be band magnetism instead of localized moments
\cite{g}. This would be consistent with the differences in the observed
saturation moment and effective moment from those expected for Ru$^{5+}$. It
is also consistent with the absence of magnetic peaks in the neutron
scattering measurements \cite{a,g}. On
the other hand, the effects of doping by annealing in oxygen appear to be
quite dependent on the Ce concentration.  For the parent composition $x=0$,
there appears to be little effect on  $T_{\mathrm{M}}$ and
$T_{\mathrm{irr}}$ \cite{a}.  For $x=0.5$, however, both  $T_{\mathrm{M}}$ and
$T_{\mathrm{irr}}$ appear to be enhanced \cite{h}.  As previously mentioned,
 $T_{\mathrm{C}}$ for this value of $x$ was significantly enhanced by
annealing in oxygen.  Thus the effects of doping either by reducing the Ce
concentration or changing the oxygen composition by annealing at high pressure
are quite different, presumably due to the fact that superconductivity resides
in the CuO$_2$ planes and the magnetic order resides in the RuO$_2$ plane.
\par
\begin{figure}[tbh]
\vspace{-10pt}
\centerline{\epsfig{file=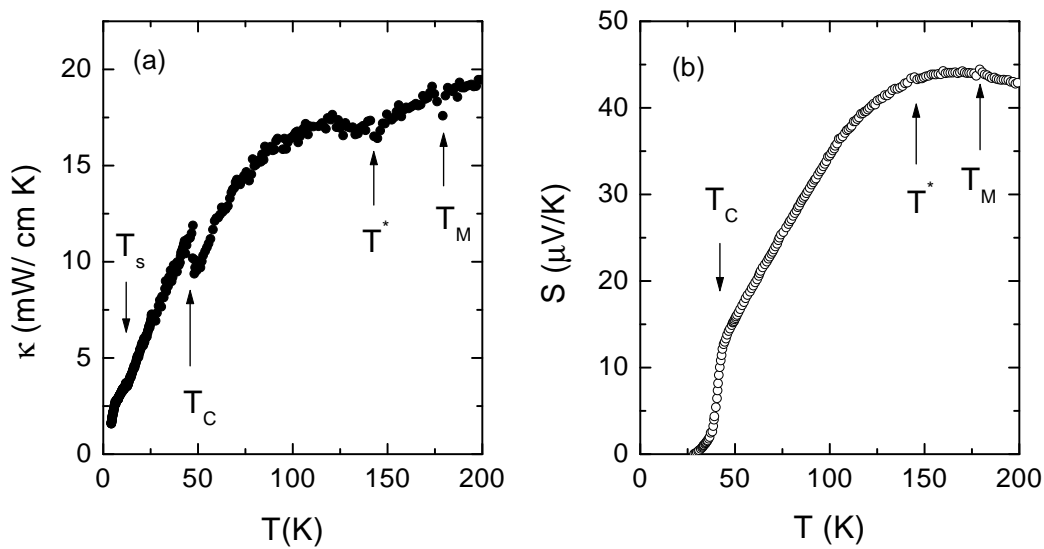,width=14cm}}
\caption{{\small {\sl  Thermal conductivity $\kappa(T)$ (a) and thermopower 
$S(T)$ (b) as a function of temperature $T$ for the same 
Eu$_{1.5}$Ce$_{0.5}$RuSr$_2$Cu$_2$O$_{10-\delta}$ sample of 
Fig. \ref{felner1}
(also taken from Ref. \cite{brian1}).  A small feature in both $S(T)$ and
$\kappa(T)$ at 180 K matches with the antiferromagnetic ordering temperature
$T_{\mathrm{M}}$, but no feature corresponding to $T_{\mathrm{irr}}$ near 77K
is seen in either $S(T)$ or $\kappa(T)$.  The superconducting transition
temperature $T_{\mathrm{C}}$ is identified as the sharp change in the slope in
$S(T)$ at 45K (coincidental with that determined from
$d\chi_{\mathrm{DC}}/dT$ in Fig. \ref{felner1}) and as the abrupt jump in
$\kappa(T)$ at 48~K. The small, reproducible feature in both $\kappa(T)$ and
$S(T)$ at $T^{\ast}=145$~K does not match with any yet reported magnetic or
structural transitions for this compound.
}}}
\label{felner2}
\end{figure}

The thermopower $S(T)$ and thermal conductivity $\kappa(T)$ for the same
sample \cite{brian1} described in Fig. \ref{felner1} is shown as a function of
temperature in Fig. \ref{felner2}.  For this sample,  $T_{\mathrm{M}}$ and
$T_{\mathrm{irr}}$ are 180 K and 77 K, respectively.  The thermopower is
positive over the entire temperature range, consistent with hole charge
carriers, and the thermal conductivity is relatively small, characteristic of
a pressed powder sample.  There is a sharp break in the slope of $S(T)$ at
$T = 45$~K, precisely the temperature indicated in Fig. \ref{felner1} for
$T_{\mathrm{C}}$.  The thermopower drops rapidly below this temperature, but
it does not go to zero until about 29 K, as might be expected for a granular
superconductor.  The behavior of the thermal
conductivity near $T_{\mathrm{C}}$ is quite unusual, in that $\kappa(T)$
increases by about 30\% in an interval of about 1 K at 48 K. Since $\kappa(T)$
and $S(T)$ were measured simultaneously, this 3~K difference between their
indicated values of $T_{\mathrm{C}}$ is probably not an experimental artifact.
Two additional small reproducible features are seen in both $S(T)$  and
$\kappa(T)$, one at $T_{\mathrm{M}} = 180$~K, which corresponds to other
indications of $T_{\mathrm{M}}$ for the sample \cite{brian1}, and one at
$T^{\ast} = 145$~K, which does not coincide with any other reported
magnetic or structural anomalies for this compound.  An additional feature in
$\kappa(T)$ is seen, a small shoulder with onset at $T_{\mathrm{s}}=13$~K.
Below about 5 K, the thermal conductivity begins to decrease rapidly.

\par
The most interesting feature is the abrupt 30\% increase in $\kappa(T)$ at
48~K.  As discussed in Sec. \ref{super} above, the thermal conductivity for an
ordinary superconductor can either increase or decrease at $T_{\mathrm{C}}$,
dependent on the dominant heat carrier and scattering mechanism.  If the
dominant carriers are phonons and the scattering mechanism is due primarily to
electrons, $\kappa(T)$ can increase below $T_{\mathrm{C}}$ as the normal
electrons are frozen out.  If the electronic component of  $\kappa$ dominates
at $T_{\mathrm{C}}$, an increase in $\kappa$  is not expected at 
$T_{\mathrm{C}}$ since the electronic part decreases below $T_{\mathrm{C}}$, 
as illustrated in Fig. \ref{figbrt}.  The freezing out of the normal 
electrons is generally very gradual because the conventional superconducting 
transition in the absence of a magnetic field is second order.  Thus, the 
narrow temperature interval ($\simeq 1$~K) for the jump is particularly 
surprising.  This behavior is quite distinct from any previously reported for
high-$T_{\mathrm{C}}$ superconductors \cite{uher}, which do, however, exhibit 
a much broader rise in $\kappa(T)$ below $T_{\mathrm{C}}$.
\par
Onset of magnetic order generally leads to a significant increase in 
$\kappa(T)$ due to loss of scattering of the electrons by spin disorder.  This 
is particulary pronounced in the data for HoNi$_2$B$_2$C, DyNi$_2$B$_2$C and 
TmNi$_2$B$_2$C (though it is probably due to a new heat carrier, spin waves, 
for the latter compound) as discussed in Sec.~\ref{boro} above (see Figs. 
\ref{dydiff}, \ref{hodiff}, and \ref{tmdiff}). In contrast, there is no strong 
signature of magnetic order in $\kappa(T)$ near the AFM ordering temperature 
$T_{\mathrm{N}}$ or the onset of weak FM at $T_{\mathrm{WF}}$ for 
ErNi$_2$B$_2$C (see Fig. \ref{erdiff}) which are both well below 
$T_{\mathrm{C}}$.  For this rutheno-cuprate sample, however, AFM order has 
already set in at $T_{\mathrm{M}} = 180$~K and weak FM at 
$T_{\mathrm{irr}}= 77$~K.   Based on the results described earlier for 
DyNi$_2$B$_2$C and TbNi$_2$B$_2$C (see Figs. \ref{dydiff} and \ref{tbdiff}), 
the magnetic ordering is expected to affect the scattering of the electrons 
primarily, with little effect on the phonon scattering.  Unfortunately, 
resistivity was not measured for this ruthen0-cuprate sample to provide an 
estimate of the electronic contribution to $\kappa(T)$ through the WF law.  
Such an estimate probably would have been very inaccurate in any case, 
however, since it is less likely that this law would be valid at the higher 
$T_{\mathrm{C}}$ values.
\par
The shoulder between 5 K and 13 K in $\kappa(T)$ is perhaps related to the 
similar "phonon" peak seen in $\kappa(T)$ for the Lu(Y)Ni$_2$B$_2$C samples 
discussed in Sec. \ref{boro} above (see Fig. \ref{yludiff}). It is quite 
probable that this peak for these borocarbides arises from the phonon channel 
as the normal electrons become less effective phonon scatterers at 
temperatures well below $T_{\mathrm{C}}$.  A similar interpretation  would 
imply a strong electron-phonon interaction in the rutheno-cuprate.  It would 
also imply a very reduced density of normal electrons in the regions where the 
magnetic order parameter was strong since the shoulder was washed out by the 
introduction of vortices with their normal cores in YNi$_2$B$_2$C through 
application of a magnetic field \cite{sera}.  The breadth of the drop to zero 
of $S(T)$ at 29 K, well below the onset of superconductivity at 45 K, may be 
the result of inhomogeneity of the doping of the sample leading to a 
percolation transition or to the Josephson junction array model suggested by 
Lorenz {\it et al.} \cite{lorenz}.  The abrupt jump in $\kappa(T)$ at 48 K 
would appear to be inconsistent either with a percolation transition or the 
Josephson junction array model, however.  Alternatively, it may result from 
vortices in the superconducting state or another, more complicated state of 
the coupled superconducting and magnetic order parameters.
\par
The thermal conductivity measurements cannot determine the detailed nature of 
the phase that leads to this abrupt jump in $\kappa$ near $T_{\mathrm{C}}$.  
The jump does strongly suggest that the transition is first order.  This may 
be due to the influence of the internal field on the superconducting state, a 
significant change in the magnetic order to accomodate coexistence of both 
types of order, or the appearance of vortices in the superconducting state to 
accommodate weak FM.  These rutheno-cuprates do appear to be quite unique in 
their behavior, however.  The fact that magnetic order first sets in at a 
temperature almost four times that of $T_{\mathrm{C}}$ is fascinating.  A 
magnetic transition at 48 K to a new magnetic state that would permit the 
superconducting phase transition at 45 K would appear to be consistent with 
the abrupt jump in  $\kappa$ and the 3 K difference in the sharp features 
near $T_{\mathrm{C}}$ seen in the simultaneous $S(T)$  and $\kappa(T)$ 
measurements.  Major efforts are needed to grow better quality and better 
characterized materials so that techniques that probe this new state on a 
microscopic scale can be more readily employed.

\section{Conclusion}
In conclusion, we have presented here an overview of the thermal conductivity
behavior in magnetic superconductors with particular emphasis on the rare 
earth nickel borocarbide intermetallic compounds and the 2122 rare earth 
rutheno-cuprates. The borocarbide family of intermetallics is of importance 
because of the availability of high quality single crystal samples and the 
large variety of phenomena - only superconductivity, coexistence of 
superconductivity and magnetic order, only magnetic order, and heavy fermion 
behavior - within the same family of compounds with identical crystal 
structure and similar electronic structure.  To complete the picture of 
thermal conductivity in this family, a description of the measurements of 
$\kappa(T)$ was given for all members for which single crystal samples were 
available, not just the four magnetic superconductors. Although only limited 
thermal conductivity data is available, the oxygen rich 
Eu$_{1.5}$Ce$_{0.5}$RuSr$_2$Cu$_{2}$O$_{10+\delta}$ compound may be unique as 
a compound that exhibits a coupled transition to weak ferromagnetism 
and superconductivity at a surprisingly high temperature ($\approx 45$~K) 
and onset of AFM near 200 K.
\par
It follows from the results presented that thermal conductivity study is 
a rather powerful technique for revealing the fundamental properties of 
these complex compounds. The technique is especially useful for study of 
phase transitions. It is clear, however, that any single method of 
investigation alone cannot give a complete picture of the physical properties
and phenomena, which are inherent in some complex compound. The thermal 
conductivity studies can, however, amplify  the data, obtained by other 
experimental methods (such as neutron diffraction, resistivity, specific 
heat, thermopower and other techniques) and give food for thoughts on 
important questions for theoretical investigations. 

\begin{acknowledgments}
	Much of the work reported here was supported in part by the Robert A. 
Welch Foundation (Grant A-0514), the Telecommunications and Informatics Task 
Force at Texas A\&M University, the Texas Center for Superconductivity and 
Advanced Materials at the University of Houston (TCSAM) and the National 
Science Foundation (Grants DMR-0103455 and DMR-0111682).
\end{acknowledgments}


\end{document}